\documentclass[a4paper,11pt]{article}
\usepackage{jheppub} 
\usepackage[utf8]{inputenc}
\usepackage{physics}
\usepackage{slashed}
\usepackage{caption}
\usepackage{xcolor}
\usepackage{comment}
\usepackage{multirow}
\usepackage{graphics}
\usepackage{float}
\usepackage{cancel}
\usepackage{soul}
\usepackage{cases}
\usepackage{array}
\usepackage{mathtools}  
\usepackage{amsfonts}
\usepackage{hyperref}
\usepackage{amsmath}
\usepackage{amssymb}
\usepackage{tcolorbox}
\usepackage{tikz}

\usepackage{amsmath,amssymb,bbm}

\usepackage[table]{xcolor}
\usepackage{tcolorbox}
\tcbuselibrary{skins, breakable}
\usepackage{float}
\newcommand{\cbar}[1]{\bar{#1}}          
\newcommand{\cund}[1]{\underline{#1}}    
\newcommand{\ctil}[1]{\tilde{#1}}        
\newcommand{\cuts}[1]{\underset{\sim}{#1}} 

\newcommand{\Block}{\mathcal{K}}

\newcommand{\Vol}{\mathrm{Vol}}
\newcommand{\nn}{\nonumber}

\newcolumntype{F}{>{\centering\arraybackslash$\displaystyle}m{3.8cm}<{$}}
\newcolumntype{R}{>{\centering\arraybackslash$\displaystyle}m{10.2cm}<{$}}

\title{The Conformal Grassmannian: A Symplectic Bi-Grassmannian for $CFT_
4$ Correlators}

\author{Aswini Bala,  Sachin Jain,  Dhruva K.S.}

\affiliation{Indian Institute of Science Education and Research,\\ Dr Homi Bhabha Road, Pashan, Pune, India}

\emailAdd{aswini.bala@students.iiserpune.ac.in}
\emailAdd{sachin.jain@iiserpune.ac.in}
\emailAdd{k.s.dhruva@students.iiserpune.ac.in}

\abstract{We introduce a formalism for conformal field theory in four dimensions: a symplectic bi-Grassmannian representation of CFT$_4$ Wightman correlators. Working in Klein space with off-shell spinor-helicity variables, we show that correlators of $\Delta = 2$ scalars and symmetric-traceless conserved currents are encoded by integrals over a pair of $n$-planes in a $2n$-dimensional symplectic vector space. These planes are constrained to be mutually symplectically orthogonal and aligned with the external kinematics. Conformal invariance, momentum conservation, and little-group covariance all follow geometrically from this structure.  We derive all two- and three-point functions involving scalars, fermions, conserved currents, and stress tensors. As a non-trivial test, we show that the construction reproduces the full set of independent conformally invariant structures of $\langle JJJ\rangle$ and $\langle TTT\rangle$ in CFT$_4$. The resulting expressions are considerably more compact than their momentum-space counterparts. They also make manifest the double copy between Yang--Mills $\langle JJJ \rangle$ and Einstein-gravity $\langle TTT \rangle$. We further present a helicity-basis reformulation that makes the GL(1,R) and SL(2,R) weights of individual helicity components explicit. This basis also provides a natural starting point for a twistor-space formulation of the correlators.}

\begin{document}

\maketitle

\section{Introduction}
Correlation functions in conformal field theory are strongly constrained by symmetry.
However, the form in which these constraints are implemented depends crucially on the
choice of variables. In position space, conformal invariance and current conservation reduce
two- and three-point functions of conserved currents and stress tensors to a finite number
of tensor structures\cite{Osborn:1993cr,Costa:2011mg}. Nevertheless, the explicit expressions rapidly become cumbersome,
especially for spinning operators and beyond three points. Momentum space correlators \cite{Maldacena:2011nz, Baumann:2020dch,Jain:2021qcl,Jain:2021vrv,Jain:2021wyn, Baumann:2021fxj,Jain:2020rmw, Baumann:2019oyu, Farrow:2018yni} provides a
natural language for applications to holography, cosmology and anomalies, but it comes
with its own complications: the conformal Ward identities become differential equations,
their solutions are expressed in terms of triple-\(K\) integrals and related special functions,
and ultraviolet singularities at coincident points require renormalisation
\cite{Bzowski:2013sza, Bzowski:2015pba, Bzowski:2017poo, Bzowski:2018fql, Coriano:2013jba}.

A different lesson has emerged from the study of scattering amplitudes. There, many
expressions that are highly complicated in Feynman-diagram variables become remarkably
simple when written in spinor-helicity, twistor or Grassmannian variables. Twistor space
makes hidden analytic and conformal structures visible \cite{Witten:2003nn}, while
Grassmannian formulations of scattering amplitudes reveal symmetries such as cyclicity, superconformal invariance and, in planar theories, Yangian symmetry in a geometric
way \cite{ArkaniHamed:2009dn, ArkaniHamed:2009sx}. The positive Grassmannian and
on-shell diagrams further show that amplitudes can often be understood as geometric
objects rather than as sums of many individual diagrams \cite{ArkaniHamed:2012nw}.{Recent work \cite{CalvoCortes:2025Suthar} has also uncovered a symplectic Grassmannian structure in massive scattering amplitudes. It was shown in the Coulomb-branch three- and four-point amplitudes of $\mathcal N=4$ SYM, to admit representations over $SpGr(n,2n)$, with the symplectic geometry arising from the special massive kinematics.}
These developments suggest that the complexity of correlation functions may also be
partly a consequence of using variables that obscure the underlying geometry.

There has been significant recent progress in applying similar ideas to conformal
correlators. In three-dimensional Lorentzian\cite{Gillioz:2019lgs, Meltzer:2021bmb, Meltzer:2020qbr} CFTs, spinor and twistor variables provide
compact expressions for Wightman functions\footnote{For more discussion on Wightman functions in momentum space CFT, look at \cite{Bautista:2019qxj}.} and make the role of helicity and conformal
Ward identities more transparent \cite{Baumann:2024fng, Jain:2023scft3, Bala:2025twistor, Bala:2025supertwistor, CarrilloGonzalez:2025twistor, Bala:2025penrose, Mazumdar:2025superpenrose, Ansari:2025twistor}. 
Relatedly, real-time Wightman
functions in AdS/CFT exhibit factorisation properties and double-copy-like structures in
appropriate kinematics \cite{Ansari:2025twistor}. More recently, Grassmannian ideas have
also appeared in the context of cosmological correlators, where an orthogonal
Grassmannian naturally encodes de Sitter/conformal constraints for massless spinning
fields \cite{Arundine:2026grassmannian,De:2026vasiliev,Bala:2026N1Grassmannian,Bala:2026N24Grassmannian,Huang:2026beyond}. These developments point toward a broader principle:
there should exist geometric variables in which conformal symmetry, little-group
covariance and the tensorial complexity of spinning correlators are simultaneously
organised.

In this paper we construct such a framework for four-dimensional conformal field
theory. Our main result is a symplectic bi-Grassmannian representation of CFT\(_4\)
Wightman correlators\footnote{The Wightman/discontinuity observables considered here are considerably simpler than
the corresponding Euclidean momentum-space correlators, while still retaining the
non-trivial tensor structures of spinning CFT correlators.
}. We work in Klein space \(\mathbb{R}^{2,2}\), where
\[
SO(2,2)\simeq SL(2,\mathbb{R})_L\times SL(2,\mathbb{R})_R ,
\]
and use an off-shell spinor-helicity parametrisation of momenta. Unlike ordinary
four-dimensional scattering amplitudes, the external momenta of CFT correlators are not
on shell. This leads to an enlarged little-group structure, \(SL(2,\mathbb{R})\times
GL(1,\mathbb{R})\), which plays a central role in our construction. Operators of spin
\(s\) are projected into little-group covariant objects, and the conformal generators take
a simple form in these variables.

The Grassmannian representation is built from a pair of matrices \(C\) and
\(\widetilde C\), each describing an \(n\)-plane inside a \(2n\)-dimensional symplectic
vector space. The two planes are constrained by symplectic orthogonality and by their
compatibility with the external spinor-helicity data. Schematically, the correlator is
represented as
\[
\Psi_n
=
\int \frac{dC}{\mathrm{Vol}(GL(n))}
     \frac{d\widetilde C}{\mathrm{Vol}(GL(n))}
\,
\delta(\widetilde C^{T}\Omega C)\,
\delta(\widetilde C^{T}\Omega\Lambda)\,
\delta(C^{T}\Omega\widetilde\Lambda)\,
\mathcal{A}_n(C,\widetilde C).
\]
The delta functions impose momentum conservation and ensure compatibility with the
external kinematics. More importantly, they make conformal invariance geometric: the
special conformal Ward identity follows directly from the symplectic orthogonality
constraint. The remaining dynamical and tensorial information is contained in
\(\mathcal{A}_n(C,\widetilde C)\), which is built from minors of \(C\) and
\(\widetilde C\) with the correct \(SL(2,\mathbb{R})\times GL(1,\mathbb{R})\)
weights.

A key feature of the formalism is that it naturally accommodates correlators with
multiple tensor structures. We reproduce all conformally invariant structures of the
three-current correlator \(\langle JJJ\rangle\) and the stress-tensor correlator
\(\langle TTT\rangle\) in four dimensions. From the AdS\(_5\) perspective, these
structures correspond respectively to Yang--Mills and higher-derivative \(F^3\)
interactions for currents, and to Einstein, Gauss--Bonnet and Weyl-cubed interactions
for stress tensors. In the Wightman/spinor-helicity representation, the relation between
the Yang--Mills current correlator and the Einstein-gravity stress-tensor correlator takes
a particularly simple double-copy form. This suggests that the symplectic
bi-Grassmannian knows not only about conformal symmetry, but also about the hidden
kinematic structures familiar from scattering amplitudes.

We also present a helicity-basis version of the construction. This reformulation makes
the \(GL(1,\mathbb{R})\) weights of individual helicity components explicit and provides
a bridge between the covariant little-group formulation and more amplitude-like
representations. It is useful both for comparing with flat-space spinor-helicity formulae
and for identifying the independent helicity structures of CFT correlators.


The paper is organised as follows- In section~\ref{sec:covariantSHvar} we develop the off-shell spinor-helicity
description of four-dimensional CFT correlators and compute several two- and three-point
Wightman functions directly. In section~\ref{sec:Covgrassframe} we introduce the covariant symplectic
bi-Grassmannian representation, prove conformal invariance, and show how the appropriate
minor structures reproduce the spinor-helicity results. In section~\ref{sec:HBgrassframework} we reformulate the
construction in a helicity basis, making the \(GL(1,\mathbb{R})\) and
\(SL(2,\mathbb{R})\) weights transparent. In section~\ref{sec:Twistor} we make the connection of the Grassmannian to the Ambitwistor/ $SL(2,\mathbb R)$ twistor. We conclude in section~\ref{sec:discussion} with a discussion of applications and future directions.

\section{Spinor Helicity Variables}\label{sec:covariantSHvar}
The aim of this section is to develop off-shell spinor helicity variables in four dimensions and apply the formalism to two and three point correlators of $\Delta=2$ scalars, spin$-\frac{1}{2}$ operators and symmetric traceless conserved currents. This is closely analogous to massive spinor-helicity variables, see for example
\cite{Arkani-Hamed:2017jhn},
with the important difference that here \(p^2\) is not fixed to a physical mass
but is treated as an arbitrary off-shell invariant.  Off-shell spinor helicity variables offer an efficient parametrization that makes both Lorentz symmetry and little-group structure transparent.
\subsection{Off-shell Spinor Helicity in $d=4$}\label{OffshellSH}
   We work in Klein space, $\mathbb{R}^{2,2}$, throughout this paper, where the spinor 
helicity variables and twistors are real-valued since 
$SO(2,2)\cong SL(2,\mathbb{R})_L\times SL(2,\mathbb{R})_R$. A general (off-shell) 
momentum vector can be traded for a pair of spinor variables as follows:
\begin{align}\label{PinSH}
p_{\alpha\dot\alpha}=\lambda_{I\alpha}\tilde\lambda^{I}_{\dot\alpha}
=\lambda_{I\alpha}\tilde\lambda_{J\dot\alpha}\,\epsilon^{IJ}.
\end{align}
Here $\lambda_{I\alpha}$ and $\tilde\lambda^{I}_{\dot\alpha}$ are real spinors transforming 
in the fundamental representations of $SL(2,\mathbb{R})_L$ and $SL(2,\mathbb{R})_R$ 
respectively, with indices $\alpha = 1,2$ and $\dot\alpha = 1,2$. The index $I=1,2$ labels 
the fundamental representation of the little group $SL(2,\mathbb{R})$, which is the subgroup 
of transformations on $(\lambda, \tilde\lambda)$ that leaves $p_{\alpha\dot\alpha}$ invariant.

This parametrization introduces an overcounting: while $p_{\alpha\dot\alpha}$ has 4 
independent real components, the pair $(\lambda_{I\alpha},\tilde\lambda^{I}_{\dot\alpha})$ 
carries $2\times 2\times 2=8$ real degrees of freedom, leaving 4 redundant directions. 
These are accounted for by two symmetries:
\begin{enumerate}
    \item $GL(1,\mathbb{R})$: The rescaling $\lambda_{I\alpha}\to r\,\lambda_{I\alpha}$,\ 
    $\tilde\lambda^{I}_{\dot\alpha}\to r^{-1}\tilde\lambda^{I}_{\dot\alpha}$ for 
    $r\in\mathbb{R}\setminus\{0\}$, contributing 1 real parameter.

    \item $SL(2,\mathbb{R})$: The transformation $\lambda_{I\alpha}\to S^{J}{}_{I}\,
    \lambda_{J\alpha}$,\ $\tilde\lambda^{I}_{\dot\alpha}\to (S^{-T})^{I}{}_{J}\,
    \tilde\lambda^{J}_{\dot\alpha}$ for $S\in SL(2,\mathbb{R})$, contributing 3 real parameters. Invariance of \eqref{PinSH} under this transformation is guaranteed since 
    the $I$-index is contracted using the $SL(2,\mathbb{R})$-invariant Levi-Civita 
    symbol $\epsilon^{IJ}$.
\end{enumerate}
Together these account for all $1+3=4$ redundant degrees of freedom, confirming that 
the full little group is $SL(2,\mathbb{R})\times GL(1,\mathbb{R})$.

While constructing correlation functions in this language, we will require polarization spinors that convert $\alpha,\Dot{\alpha}$ into the little group indices.
\begin{align}\label{polarizationspinors}
    \zeta_{\alpha}^{I}=\frac{\lambda_\alpha^I}{\sqrt{\text{Det}(\lambda)}},\tilde{\zeta}_{\Dot{\alpha}}^{I}=\frac{\tilde{\lambda}_{\Dot{\alpha}}^I}{\sqrt{\text{Det}(\tilde\lambda)}}.
\end{align}
These are chosen such that they project Lorentz indices into little-group indices while remaining invariant \footnote{Let us note that the choice of normalization in Eq\eqref{polarizationspinors} might appear to break $GL(1,\mathbb{R})$ to $GL(1,\mathbb{R})/\mathbb{Z}_2$, however a compensating normalization for the current operator in Eq\eqref{Jsrescaled} together with the polarization rescaling restores the $GL(1,\mathbb{R})$.     } under the $GL(1,\mathbb{R})$ redundancy.
Given spin$-\frac{1}{2}$ operators we construct,
\begin{align}\label{spinhalfI}
    O_{\frac{1}{2}}^{I}(\lambda,\tilde{\lambda})=\zeta_\alpha^I O_{\frac{1}{2}}^{\alpha}(p^\mu),~\tilde{O}_{\frac{1}{2}}^{I}(\lambda,\tilde{\lambda})=\tilde{\zeta}_{\Dot{\alpha}}^I \tilde{O}_{\frac{1}{2}}^{\Dot{\alpha}}(p^\mu).
\end{align}
To form the polarizations of symmetric traceless conserved currents, we can take symmetrized products of these quantities. For example, consider a spin$-1$ current. We have,
\begin{align}\label{spin1IJ}
    J^{(IJ)}(\lambda,\tilde{\lambda})=\zeta_{\alpha}^{(I}\tilde{\zeta}_{\Dot{\alpha}}^{J)} J^{\alpha\Dot{\alpha}}(p^\mu),
\end{align}
and similarly for higher spin\footnote{Our convention for symmetrization is,
\begin{align}
    &J^{(IJ)}=\frac{1}{2!}(J^{IJ}+J^{JI}),\notag\\
    &T^{(IJKL)}=\frac{1}{4!}(T^{IJKL}+\cdots T^{LKJI}),
\end{align}
and so on.}

\begin{align}\label{spinsIJ}
    J_s^{(I_1,\cdots,I_m,J_1,\cdots, J_{n})}(\lambda,\tilde{\lambda})=\zeta_{\alpha_1}^{(I_1}\cdots \zeta_{\alpha_m}^{I_m}\tilde{\zeta}_{\Dot{\alpha}_1}^{J_1}\cdots\tilde{\zeta}_{\Dot{\alpha}_n}^{J_n)}J_s^{\alpha_1,\cdots,\alpha_m,\Dot{\alpha}_1,\cdots,\Dot{\alpha}_n}(p^\mu),~~m+n=2s.
\end{align}
Let us note that for currents with integer spin $m=n=s$ whereas for half integer spin currents $|m-n|=1.$
Finally, we will find it useful to normalize these operators so that the conformal generators act simply on them. In particular, we have,
\begin{align}\label{Jsrescaled}
    \hat{J}^{(I_1,\cdots,I_m,J_1,\cdots J_{n})}_s(\lambda,\tilde{\lambda})=\frac{J_s^{(I_1,\cdots,I_m,J_1,\cdots, J_{n})}(\lambda,\tilde{\lambda})}{\text{Det}(\lambda)^{\frac{m}{2}}\text{Det}(\tilde{\lambda})^{\frac{n}{2}}},~~m+n=2 s.
\end{align}
Thus, under the $SL(2,\mathbb{R}) \times GL(1,\mathbb{R})$ little group we have,
\begin{align}\label{littlegroupscaling}
    \hat{J}^{(I_1,\cdots,I_m,J_1,\cdots J_{n})}_s(\lambda,\tilde{\lambda})\to \frac{1}{r^{m-n}}\big(S^{I_1}_{I_1'}\cdots S^{I_m}_{I_m'}\big)\big(S^{J_1}_{J_1'}\cdots S^{J_n}_{J_n'}\big)\hat{J}^{(I_1',\cdots,I_m',J_1',\cdots J_{n}')}_s(\lambda,\tilde{\lambda})
\end{align}
$r\in\mathbb{R}/\{0\}$ is the scaling parameter corresponding to $GL(1,\mathbb{R})$ and $S_I^{J}$ are the $SL(2,\mathbb{R})$ matrices.
For the special case of $m=n$ (integer-spins), we have invariance under the $GL(1,\mathbb{R})$ subgroup of the little group. However, half-integer spins for which $|m-n|=1$, has a non-trivial scaling behavior that we must take into account. 
\subsection{The Observables of interest}
The objects of interest to us are correlation functions of these currents (we suppress indices below) 
\begin{align}
    \langle J_{s_1}\cdots J_{s_n}\rangle.
\end{align}
In \cite{Bzowski:2017poo}, momentum-space correlation functions involving scalar and spinning operators in $d=4$ were shown to be generically divergent, necessitating renormalization. However, in light of recent developments \cite{Bautista:2019qxj}, focusing instead on the discontinuities of these correlators—closely related to Wightman functions—yields results that are finite and free from divergences.
For example, we take a $\Delta=2$ scalar operator. Its two point function is,
\begin{align}
    \langle\langle O_2(-p)O_2(p)\rangle\rangle=\log(\frac{p^2}{\mu^2}).
\end{align} 
We now take a discontinuity about $p^2=0$. The result is a constant which is divergence free \footnote{To obtain the Wightman function, all one needs to do is multiply this constant by the Heaviside theta functions that enforce the spectral conditions and characterize operator ordering. For example we find,
\begin{align}
    \theta(-p^2)\theta(p^0)\text{Disc}_{p^2}\langle\langle O_2(-p)O_2(p)\rangle\rangle=\theta(-p^2)\theta(p^0)\pi i,
\end{align}
which is the correct Wightman function up to a proportionality constant.}. A similar feature appears for current two-point functions, which are also UV divergent. Nevertheless, their discontinuities—and therefore the corresponding Wightman functions—are finite. For three-point functions, we thus concentrate on the discontinuities with respect to $p_1^2$ and $p_3^2$, which correspond to Wightman correlators in which the middle operator carries space-like momentum\footnote{Throughout, we work with Wightman functions with the spectral theta functions stripped off, since they can be reinstated unambiguously.}
\cite{Ansari:2025twistor}. 

With that discussion at hand, let us now state the conformal Ward identities in this formalism. We have, apart from the $GL(2,\mathbb{R})$ covariance that constrains each operator \eqref{littlegroupscaling}, the set of $15$ conformal Ward identities.
\begin{align}\label{WardId}
    \sum_{i=1}^{n}\mathcal{L}_i\langle J_{s_1}\cdots J_{s_n}\rangle=0,
\end{align}
where $\mathcal{L}_i\in\{P_i^\mu,D,M_i^{\mu\nu},K^\mu\}$. We have,
\begin{align}
    P^{\alpha\Dot{\alpha}}=\lambda_I^{\alpha}\tilde{\lambda}^{I\Dot{\alpha}}, K_{\alpha\Dot{\alpha}}=\frac{\partial^2}{\partial\lambda_I^{\alpha}\partial\tilde{\lambda}^{I\Dot{\alpha}}}.
\end{align}
With the formalism being set, we can now turn to examples of correlation functions in these variables. Before we present the results, the notation given below will prove useful in what follows:
\begin{align}
    \langle 1^I 2^J\rangle=\lambda_{1\alpha}^{I}\lambda_{2}^{J\alpha},~~[1^I 2^J]=\tilde{\lambda}_{1\alpha}^{I}\tilde{\lambda}_{2}^{J\alpha}.
\end{align}
The squared magnitude of the momenta is,
\begin{align}
    p^2=\text{Det}(\lambda)\text{Det}(\tilde{\lambda}).
\end{align}
Indices are raised and lowered using the Levi-Civita symbols. For example,
\begin{align}
    \lambda^{I \alpha}=\epsilon^{IJ}\epsilon^{\alpha\beta}\lambda_{J\beta},~~\tilde{\lambda}^{I \Dot{\alpha}}=\epsilon^{\Dot{\alpha}\Dot{\beta}}\tilde{\lambda}^{I}_{\Dot{\beta}}.
\end{align}
\subsection{Two point functions}\label{sec:2pointSH}
We begin with three distinct examples of two point functions and then generalize to arbitrary spin. This will illustrate the utility of our spinor helicity formalism.
\subsubsection{$\langle O_2O_2\rangle$}
For scalar operators, the two point function (stripped Wightman function) is simply a constant
\begin{align}\label{O2O2SH}
    \langle O_2(p_1)O_2(p_2)\rangle=(2\pi)^4\delta^4(p_1+p_2).
\end{align}
\subsubsection{$\langle O_{\frac{1}{2}}^{I}\tilde{O}_{\frac{1}{2}}^{J}\rangle$}
In momentum space we have,
\begin{align}
    \langle O_{\frac{1}{2}}^{\alpha}(p_1)\tilde{O}_{\frac{1}{2}}^{\Dot{\alpha}}(p_2)\rangle=p_1^{\alpha\Dot{\alpha}}(2\pi)^4 \delta^4(p_1+p_2).
\end{align}
Following the steps outlined to construct \eqref{Jsrescaled} we find,
\begin{align}\label{OhOhSH}
    \langle \hat{O}_{\frac{1}{2}}^{I}(\lambda_1,\tilde{\lambda}_1)\hat{\tilde{O}}_{\frac{1}{2}}^{J}(\lambda_2,\tilde{\lambda}_2)\rangle&=\frac{1}{\sqrt{\text{Det}(\lambda_1)\text{Det}(\tilde{\lambda}_2)}}\frac{\lambda_{1\alpha}^I\tilde{\lambda}_{2\Dot{\alpha}}^J(\lambda_{1K}^{\alpha}\tilde{\lambda}_{1}^{K\Dot{\alpha}})}{\sqrt{\text{Det}(\lambda_1)\text{Det}(\tilde{\lambda}_2)}}(2\pi)^4\delta^4(p_1+p_2)\notag\\
    &=-\frac{[ 1^I 2^J]}{\text{Det}({\tilde\lambda}_2)}(2\pi)^4\delta^4(p_1+p_2).
\end{align}
Note that under the $GL(1)_1\times GL(1)_2$ little group scaling, each operator transforms in perfect accordance with \eqref{littlegroupscaling}.

\subsubsection{$\langle J^{I_1J_1}J^{I_2J_2}\rangle$}
We consider the more complicated example of a spin$-1$ conserved current. Its stripped Wightman function in momentum space is given by,
\begin{align}
    \langle J^{\alpha\Dot{\alpha}}(p_1)J^{\beta\Dot{\beta}}(p_2)\rangle=(\sigma_\mu)^{\alpha\Dot{\alpha}}(\sigma_\nu)^{\beta\Dot{\beta}}\bigg(p_1^2\pi^{\mu\nu}(p_1)(2\pi)^4\delta^4(p_1+p_2)\bigg),
\end{align}
where $\pi^{\mu\nu}(p)=\eta^{\mu\nu}-\frac{p^\mu p^\nu}{p^2}$ is the transverse projector. After using the fact that $(\sigma^\mu)^{\alpha\Dot{\alpha}}(\sigma_\mu)^{\beta\Dot{\beta}}=-2\epsilon^{\alpha\beta}\epsilon^{\Dot{\alpha}\Dot{\beta}}$, constructing \eqref{Jsrescaled} and using the following identity that follows from momentum conservation ,
\begin{align}
    [1^I 2^J]=-\frac{\text{Det}(\tilde{\lambda}_2)}{\text{Det}(\lambda_1)}\langle 1^I 2^J\rangle,
\end{align}
we obtain,
\begin{align}\label{JJSH}
    \langle \hat{J}^{(I_1J_1)}(\lambda_1,\tilde{\lambda}_1)\hat{J}^{(I_2J_2)}(\lambda_2,\tilde{\lambda}_2)\rangle&=-2(2\pi)^4\delta^4(p_1+p_2)\frac{\langle 1^{I_1}2^{I_2}\rangle[1^{I_2} 2^{J_2}]}{p_1^2}\notag\\
    &=2(2\pi)^4\delta^4(p_1+p_2)\frac{\langle 1^{I_1}2^{I_2}\rangle\langle 1^{J_1}2^{J_2}\rangle}{\text{Det}(\lambda_1)\text{Det}(\lambda_2)},
\end{align}
with $(I_1,I_2)$ and $(J_1,J_2)$ symmetrized. Having understood the general pattern we can now generalize this result to arbitrary spins.
\subsubsection{$\langle J_s J_s\rangle$}
Based on the pattern we have seen so far, a natural ansatz is,
\begin{align}\label{JsJsSH}
    \langle \hat{J}^{(I_1\cdots I_{2s})}(\lambda_1,\tilde{\lambda}_1)\hat{J}^{(J_1\cdots J_{2s})}(\lambda_2,\tilde{\lambda}_2)\rangle&=(2\pi)^4\delta^4(p_1+p_2)\frac{\langle 1^{I_1}2^{J_1}\rangle\langle 1^{I_2}2^{J_2}\rangle\cdots\langle 1^{I_{2s}}2^{J_{2s}}\rangle}{\text{Det}(\lambda_1)^{2m}\text{Det}(\lambda_2)^{2m}}.
\end{align}
where $(I_1,\cdots,I_{2s})$ and $(J_1,\cdots,J_{2s})$ are totally symmetrized. This quantity is conformally invariant \eqref{WardId} and is consistent with the little group property \eqref{littlegroupscaling}.

We now proceed to the three point level.
\subsection{Three point functions}\label{sec:3pointSH}
Three-point functions in \( d=4 \) are, in general, structurally more involved due to the presence of branch cuts in their dependence on momentum magnitudes \cite{Bzowski:2015pba,Bzowski:2017poo,Bzowski:2018fql}, which complicates their analytic structure. However,  Wightman functions with \( p_1^\mu, p_3^\mu \) time-like and \( p_2^\mu \) space-like, which  can be obtained by taking discontinuities of the Euclidean correlators with respect to \( p_1^2 \) and \( p_3^2 \) turns out to be quite simple and elegant. In principle, one can therefore take the known results of \cite{Bzowski:2015pba,Bzowski:2017poo,Bzowski:2018fql} and extract the desired Wightman functions via appropriate discontinuities.  However, these expressions are typically quite involved, as they contain polylogarithmic structures, see \cite{Bzowski:2017poo}, making the analytic continuation non-trivial. It is therefore more efficient to implement the discontinuity directly at the level of the Triple-K integrand and then perform the integrals.
\subsubsection{$\langle O_2O_2O_2\rangle$}
Following the formalism of \cite{Ansari:2025twistor}, using the equation of motion to compute this $\Delta=2$ three point Wightman function with $p_2^2>0$ results in,
\begin{align}
    &\langle 0|O_2(p_1)O_2(p_2)O_2(p_3)|0\rangle\theta(p_2^2)\notag\\&=\theta(-p_1^2)\theta(p_2^2)\theta(-p_3^2)\theta(-p_1^0)\theta(p_3^0)\int_0^\infty dz~z~ J_{0}(|p_1| z)K_{0}(|p_2|z)J_0(|p_3|z).
\end{align}
One can also check that this quantity matches the result obtained via analytic continuation \cite{Bautista:2019qxj}. This quantity is obtained by taking a discontinuity with respect to $p_1^2$ and $p_3^2$ of the usual Triple-K integral $I_{1\{0,0,0\}}$ \cite{Bzowski:2013sza}.  The stripped Wightman function is given by
\begin{align}
    \langle O_2(p_1)O_2(p_2)O_2(p_3)\rangle&=\int_0^\infty dz~z~ J_{0}(|p_1| z)K_{0}(|p_2|z)J_0(|p_3|z)\nonumber\\
    &= 
\frac{1}{
\sqrt{
\left(|p_2|^2+(|p_1|+|p_3|)^2\right)
\left(|p_2|^2+(|p_1|-|p_3|)^2\right)
}
}
\end{align}
where the details of the integration is given in the Appendix \ref{Bessel}.

As explained earlier, we are interested in the configuration where the first and third operators carry time-like momenta, while the middle operator carries space-like momentum. This corresponds to the analytic continuation
\begin{equation}
|p_1| = -i p_1, \qquad |p_2| = p_2, \qquad |p_3| = +i p_3.
\end{equation}

Substituting this, we obtain
\begin{align}
\langle O_2(p_1) O_2(p_2) O_2(p_3) \rangle
&=
\frac{1}{
\sqrt{
-(p_1 + p_2 + p_3)(-p_1 + p_2 + p_3)(p_1 - p_2 + p_3)(p_1 + p_2 - p_3)
}
}\nonumber\\
&=
\frac{1}{\sqrt{-\mathcal{J}^2}}.
\end{align}

Here we have defined
\begin{equation}
\mathcal{J}^2
=
(p_1 + p_2 + p_3)(p_1 + p_2 - p_3)(p_1 - p_2 + p_3)(-p_1 + p_2 + p_3),
\end{equation}
which corresponds to Heron's formula for the area of a triangle.

\subsubsection{$\langle J^{IJ}O_2O_2\rangle$}
The Euclidean correlator in momentum space is given by \cite{Bzowski:2018fql},
\begin{align}
    \langle J^{\mu}(p_1)O_2(p_2)O_2(p_3)\rangle_E=\pi^{\mu}_{\nu}(p_1)p_2^{\nu}p_1\frac{\partial}{\partial p_1}I_{1,\{000\}},
\end{align}
where $I_{1,\{000\}}$ is the triple-K integral,
\begin{align}
    I_{1,\{000\}}=\int_0^\infty dz z K_0(p_1 z)K_0(p_2 z)K_0(p_3 z).
\end{align}
Following same procedure as for the scalar case, the Wightman function can be written as 
\begin{align}
    \langle J^{\mu}(p_1)O_2(p_2)O_2(p_3)\rangle=\pi^{\mu}_{\nu}(p_1)p_2^{\nu}p_1\frac{\partial}{\partial p_1}\int_0^\infty dz z J_0(|p_1| z)K_0(|p_2| z)J_0(|p_3| z),
\end{align}
which is an integral we have already evaluated for the scalar three point function. Contracting the free index of the current with the transverse polarization vector $\epsilon^{IJ}_{\mu}$ and rescaling \eqref{Jsrescaled} results in,
\begin{align}\label{JOOSH}
    \langle \hat{J}^{IJ}(p_1)\hat{O}_2(p_2)\hat{O}_2(p_3)\rangle&=(\epsilon^{IJ}_1\cdot p_2)\frac{p_1(-p_1^2+p_2^2+p_3^2)}{(-\mathcal{J}^2)^{\frac{3}{2}}}\notag\\
    &=\frac{\langle 1^{(I}2_{K}\rangle[1^{J)} 2^K](-p_1^2+p_2^2+p_3^2)}{(-\mathcal{J}^2)^{\frac{3}{2}}}.
\end{align}
\subsubsection{$\langle T^{(IJKL)}O_2O_2\rangle$}
We proceed to an example involving a stress tensor now. The momentum space correlator is given in \cite{Bzowski:2018fql}. Taking its discontinuity with respect to $p_1^2$ and $p_3^2$ to obtain the Wightman function results in,
\begin{align}
    \epsilon_{1\mu}\epsilon_{1\nu}\langle T^{\mu\nu}(p_1)O_2(p_2)O_2(p_3)\rangle&=(\epsilon_1\cdot p_2)^2 p_1\bigg(p_1\frac{\partial}{\partial p_1}-1\bigg)\frac{\partial}{\partial p_1}\frac{1}{\sqrt{-\mathcal{J}^2}}\notag\\
    &=(\epsilon_1\cdot p_2)^2\frac{8p_1^4(p_1^4+p_2^4+p_3^4+4p_2^2p_3^2-2p_1^2p_2^2-2p_1^2p_3^2))}{(-\mathcal{J}^2)^{\frac{5}{2}}}.
\end{align}
Plugging in the spinor helicity expressions and simplifying then results in,
\begin{align}\label{TOOSH}
    \langle T^{I_1J_1K_1L_1}O_2O_2\rangle=\frac{\langle 1^{I_1}2^I\rangle[1^{J_1}2_I]\langle 1^{K_1}2^J\rangle[1^{L_1}2_J](2p_2^2p_3^2+4(p_2\cdot p_3)^2}{(-\mathcal{J}^2)^{\frac{5}{2}}}.
\end{align}

\subsubsection{$\langle O_{\frac{1}{2}}^I\tilde{O}_{\frac{1}{2}}^JO_2\rangle$}
To compute this correlator, we consider the free supersymmetric theory involving a massless Weyl fermion and a complex scalar. In the free theory realizations we have $O_{\frac{1}{2}}^{\alpha}=\psi^\alpha\Bar{\phi}$ and $\tilde{O}_{\frac{1}{2}}^{\Dot{\alpha}}=(\psi^\dagger)^{\Dot{\alpha}}\phi$ and $O_2=\Bar{\phi}\phi$. Performing Wick contractions results in the following expression for the Euclidean correlator.
\begin{align}
    \langle O_{\frac{1}{2}}^{\alpha}(p_1)\tilde{O}_{\frac{1}{2}}^{\Dot{\alpha}}(p_2)O_2(p_3)\rangle_E&=\int \frac{d^4 l~l^{\alpha\Dot{\alpha}}}{l^2(l-p_1)^2(l+p_2)^2}\notag\\
    &=\frac{1}{8\pi^2}\bigg(-p_1^{\alpha\Dot{\alpha}}p_2\frac{\partial}{\partial p_2}I_{1\{000\}}+p_2^{\alpha\Dot{\alpha}}p_1\frac{\partial}{\partial p_1}I_{1\{000\}}\bigg).
\end{align}
The corresponding Wightman function found by taking the discontinuity with respect to $p_1^2,p_3^2$ is thus given by,
\begin{align}
     \langle O_{\frac{1}{2}}^{\alpha}(p_1)\tilde{O}_{\frac{1}{2}}^{\Dot{\alpha}}(p_2)O_2(p_3)\rangle=\frac{1}{(-\mathcal{J}^2)^{\frac{3}{2}}}\bigg(p_1^{\alpha\Dot{\alpha}}p_2^2(p_1^2-p_2^2+p_3^2)-p_2^{\alpha\Dot{\alpha}}p_1^2(-p_1^2+p_2^2+p_3^2)\bigg),
\end{align}
where we used the fact that $\text{Disc}_{p_1^2}\text{Disc}_{p_3^2}I_{1\{000\}}\sim \frac{1}{(-\mathcal{J}^2)^{\frac{1}{2}}}$. To obtain the spinor helicity expression, we now contract with the polarization spinors \eqref{spinhalfI} and rescaling \eqref{Jsrescaled} results in,
\begin{align}\label{OhOhO2SH}
    \langle \hat{O}_{\frac{1}{2}}^{I}\tilde{\hat{O}}_{\frac{1}{2}}^J\hat{O}_2\rangle=\frac{1}{(-\mathcal{J}^2)^{\frac{3}{2}}}\bigg([1^I 2^J]\text{Det}(\lambda_2)(p_1^2-p_2^2+p_3^2)-\langle 1^I 2^J\rangle\text{Det}(\tilde{\lambda}_1)(-p_1^2+p_2^2+p_3^2)\bigg).
\end{align}

\subsubsection{$\langle J^{I_1J_1}J^{I_2J_2}O_2\rangle$}
The Euclidean $\langle JJO_2\rangle_E$ correlator can be found in \cite{Bzowski:2018fql}. 
\begin{align}
    \langle J(p_1,\epsilon_1)J(p_2,\epsilon_2)O_2(p_3)\rangle=(\epsilon_1\cdot p_2)(\epsilon_2\cdot p_3)A_1+(\epsilon_1\cdot\epsilon_2)A_2,
\end{align}
where,
\begin{align}
    &A_1=p_1 p_2\frac{\partial^2}{\partial p_1\partial p_2}I_{1,\{000\}},\notag\\
    &A_2=I_{2\{111\}}^{\text{fin}}+\frac{1}{6}\text{log}\big(\frac{p_3^4}{p_1^2p_2^2}\big)+\frac{1}{2}.
\end{align}
Taking a discontinuity with respect to $p_1^2,p_3^2$ and converting to spinor helicity results in,
\begin{align}\label{JJO2SH}
    \langle \hat{J}^{I_1J_1}\hat{J}^{I_2J_2}\hat{O}_2\rangle&=\frac{1}{(-\mathcal{J}^2)^{\frac{5}{2}}}\bigg(-2p_3^2(p_1^4+(p_2^2-p_3^2)^2-2p_1^2(p_2^2+p_3^2))\langle 1^{I_1}2^{I_2}\rangle[1^{J_1}2^{J_2}]\notag\\
    &+2\langle 1^{I_1}2_I\rangle[1^{J_1}2^I]\langle 2^{I_2}1_J\rangle[2^{J_2}1^J]((p_1^2-p_2^2)^2+(p_1^2+p_2^2)p_3^2-2p_3^4)\bigg).
\end{align}
So far, all the correlators we have considered have a unique structure determined by conformal symmetry. The remaining examples on the other hand that we now turn to, have multiple possible structures consistent with conformal invariance.
\subsubsection{$\langle J^{I_1J_1}J^{I_2J_2}J^{I_3J_3}\rangle$}
Consider a correlator of three spin$-1$ currents valued in the adjoint representation of a global symmetry group such as $SU(N)$. Conformal invariance allows for the existence of two independent structures. From the AdS$_5$ bulk perspective, the first of these corresponds to the usual three point interaction arising from the Yang-Mills action of the dual gluons. The other is a contribution from a higher derivative $\text{tr}(F^3)$ term. In \cite{Bzowski:2017poo}, these are respectively the coefficient of $C_{JJ}$ and $C_1$. Computing the Wightman functions by taking a discontinuity with respect to $p_1^2,p_3^2$ for the Yang-Mills case results in,
\begin{align}\label{JJJYM}
    \langle J(p_1,\epsilon_1)J(p_2,\epsilon_2)J(p_3,\epsilon_3)\rangle_{YM}=\frac{\bigg((\epsilon_1\cdot\epsilon_2)(\epsilon_3\cdot p_1)+(\epsilon_1\cdot \epsilon_3)(\epsilon_2\cdot p_3)+(\epsilon_2\cdot\epsilon_3)(\epsilon_1\cdot p_2)\bigg)p_1^2 p_2^2 p_3^2}{(-\mathcal{J}^2)^{\frac{3}{2}}}.
\end{align}
In the $F^3$ case we find,
\footnotesize
\begin{align}\label{JJJF3}
     &\langle J(p_1,\epsilon_1)J(p_2,\epsilon_2)J(p_3,\epsilon_3)\rangle_{F^3}=p_1p_2p_3\bigg(-\frac{\partial^3}{\partial p_1\partial p_2\partial p_3}\big(\frac{1}{\sqrt{-\mathcal{J}^2}}\big (\epsilon_1\cdot p_2)(\epsilon_2\cdot p_3)(\epsilon_3\cdot p_1)\notag\\&+p_3\frac{\partial^2}{\partial p_1\partial p_2}\big(\frac{1}{\sqrt{-\mathcal{J}^2}}\big)(\epsilon_1\cdot \epsilon_2)(\epsilon_3\cdot p_1)+p_2\frac{\partial^2}{\partial p_1\partial p_3}\big(\frac{1}{\sqrt{-\mathcal{J}^2}}\big)(\epsilon_1\cdot \epsilon_3)(\epsilon_2\cdot p_3)+p_1\frac{\partial^2}{\partial p_2\partial p_3}\big(\frac{1}{\sqrt{-\mathcal{J}^2}}\big)(\epsilon_2\cdot \epsilon_3)(\epsilon_1\cdot p_2)\bigg).
\end{align}
\normalsize
The spinor helicity expressions can be found using the simple substitutions,
\begin{align}\label{SHdotprods}
    \epsilon_i^{I_i J_i}\cdot\epsilon_{j}^{I_j J_j}=-\frac{1}{2p_ip_j}\langle i^{I_i}j^{J_i}\rangle[i^{J_i}j^{J_j}]~,\epsilon_{i}^{I_i J_i}\cdot p_j=-\frac{1}{2p_i}\langle i^{I_i}j_I\rangle[i^{J_i}j^I].
\end{align}

\subsubsection{$\langle T^{I_1J_1K_1L_1}J^{I_2J_2}J^{I_3J_3}\rangle$}
Starting with the expression for the Euclidean correlator in \cite{Bzowski:2017poo}, we take a discontinuity with respect to $p_1^2$ and $p_3^2$. This yields,
\begin{align}\label{TJJSH}
    \langle &T(p_1,\epsilon_1)J(p_2,\epsilon_2)J(p_3,\epsilon_3)=(\epsilon_1\cdot p_2)^2(\epsilon_2\cdot p_3)(\epsilon_3\cdot p_1)A_1(p_1,p_2,p_3)\notag\\
    &+(\epsilon_2\cdot\epsilon_3)(\epsilon_1\cdot p_2)^2A_2(p_1,p_2,p_3)+(\epsilon_1\cdot\epsilon_2)(\epsilon_1\cdot p_2)(\epsilon_3\cdot p_1)A_3(p_1,p_2,p_3)\notag\\
    &+(\epsilon_1\cdot\epsilon_3)(\epsilon_1\cdot p_2)(\epsilon_2\cdot p_3)A_3(p_1,p_3,p_2)+(\epsilon_1\cdot\epsilon
    _3)(\epsilon_1\cdot\epsilon_2)A_4(p_1,p_2,p_3),
\end{align}
where,
\begin{align}
    &A_1=C_1 p_1p_2p_3\bigg(p_1\frac{\partial}{\partial p_1}-1\bigg)\tilde{I}_{1,\{000\}},\notag\\
    &A_2=2C_J\big(2-p_1\frac{\partial}{\partial p_1}\big)\tilde{I}_{2,\{111\}}-C_1 p_1^3p_2p_3\frac{\partial^3}{\partial p_1\partial p_2\partial p_3}\tilde{I}_{1\{000\}},\notag\\
    &A_3=2C_J\big(2-p_1\frac{\partial}{\partial p_1}\big)\tilde{I}_{2,\{111\}}+2C_1p_1p_2p_3^2\bigg(\frac{\partial^2}{\partial p_1\partial p_2}\tilde{I}_{1,\{000\}}-p_1\frac{\partial^3}{\partial p_1\partial p_2\partial p_1}\tilde{I}_{1,\{000\}}\bigg),\notag\\
    &A_4=-2C_Jp_1^2\tilde{I}_{2,\{111\}}-2C_1p_2^2p_3^2p_1\big(1-p_1\frac{\partial}{\partial p_1}\big)\frac{\partial}{\partial p_1}\tilde{I}_{1,\{000\}},
\end{align}
where,
\begin{align}
    &\tilde{I}_{2,\{111\}}=-\frac{4p_1^2p_2^2p_3^2}{\mathcal{J}^2}\tilde{I}_{1,\{000\}},\notag\\
    &\tilde{I}_{1,\{000\}}=\frac{1}{\sqrt{-\mathcal{J}^2}}.
\end{align}
The corresponding spinor helicity expression can be found using \eqref{SHdotprods}.
\subsubsection{$\langle T^{I_1J_1K_1L_1}T^{I_2J_2K_2L_2}T^{I_3J_3K_3L_3}\rangle$}
Finally, we consider the three point function of the stress tensor. In $d=4$, there are three independent conformally invariant structures. From the bulk AdS$_5$ perspective, these correspond to three point graviton interactions arising from the Einstein-Hilbert term, the Gauss-Bonnet term (which is not topological in $d\ge 5$) and a $\text{Weyl}^3$ curvature correction. Similar to the previous examples, we can compute the corresponding Wightman functions by taking a discontinuity with respect to $p_1^2,p_3^2$ of the results given in \cite{Bzowski:2017poo}. The result is,
\begin{align}\label{TTTSH}
    &\langle T(p_1,\epsilon_1)T(p_2,\epsilon_2)T(p_3,\epsilon_3)\rangle=(\epsilon_1\cdot p_2)^2(\epsilon_2\cdot p_3)^2(\epsilon_3\cdot p_1)^2A_1(p_1,p_2,p_3)\notag\\
    &+(\epsilon_1\cdot\epsilon_2)(\epsilon_1\cdot p_2)(\epsilon_2\cdot p_3)(\epsilon_3\cdot p_1)^2A_2(p_1,p_2,p_3)+(\epsilon_2\cdot\epsilon_3)(\epsilon_1\cdot p_2)^2(\epsilon_2\cdot p_3)(\epsilon_3\cdot p_1)A_2(p_3,p_2,p_1)\notag\\
    &+(\epsilon_1\cdot\epsilon_3)(\epsilon_1\cdot p_2)(\epsilon_2\cdot p_3)^2(\epsilon_3\cdot p_1)A_2(p_1,p_3,p_2)+(\epsilon_1\cdot\epsilon_2)^2(\epsilon_3\cdot p_1)^2A_3(p_1,p_2,p_3)\notag\\
    &+(\epsilon_2\cdot \epsilon_3)^2(\epsilon_1\cdot p_2)^2A_3(p_3,p_2,p_1)+(\epsilon_1\cdot \epsilon_3)^2(\epsilon_2\cdot p_3)^2 A_3(p_1,p_3,p_2)\notag\\
    &+(\epsilon_1\cdot\epsilon_3)(\epsilon_2\cdot\epsilon_3)(\epsilon_1\cdot p_2)(\epsilon_2\cdot p_3)A_4(p_1.p_2,p_3)+(\epsilon_1\cdot\epsilon_3)(\epsilon_1\cdot\epsilon_2)(\epsilon_2\cdot p_3)(\epsilon_3\cdot p_1)A_4(p_3,p_2,p_1)\notag\\
    &+(\epsilon_1\cdot\epsilon_2)(\epsilon_2\cdot \epsilon_3)(\epsilon_1\cdot p_2)(\epsilon_3\cdot p_1)A_4(p_1,p_3,p_2)+(\epsilon_1\cdot\epsilon_2)(\epsilon_2\cdot\epsilon_3)(\epsilon_1\cdot\epsilon_3)A_5(p_1,p_2,p_3),
\end{align}
where the form factors (which are the discontinuities of the ones in \cite{Bzowski:2017poo}) are given by,
\begin{align}
    &A_1=C_1 \tilde{I}_{7,\{222\}},\notag\\
    &A_2=2(a+c-2C_1p_3\frac{\partial}{\partial p_3})\tilde{I}_{5,\{222\}},\notag\\
    &A_3=2\big(2c-(a+c+C_1)p_3\frac{\partial}{\partial p_3}+C_1p_3^2\frac{\partial^2}{\partial p_3^2}\big)\tilde{I}_{3,\{222\}},\notag\\
    &A_4=4(c-a+(a+c)p_3\frac{\partial}{\partial p_3}+2C_1(8-4\sum_{j=1}^{3}p_j\frac{\partial}{\partial p_j}+p_1 p_2\frac{\partial^2}{\partial p_1 \partial p_2}\big)\tilde{I}_{3,\{222\}},\notag\\
    &A_5=2(a+c)\big(32-8\sum_{j=1}^{3}p_j\frac{\partial}{\partial p_j}+2\sum_{i<j}p_ip_j\frac{\partial^2}{\partial p_i\partial p_j}\big)\tilde{I}_{1,\{222\}}-8 C_1 p_1^3 p_2^3 p_3^3\frac{\partial^3}{\partial p_1\partial p_2\partial p_3}\tilde{I}_{1,\{000\}},
\end{align}
where,
\begin{align}
    &\tilde{I}_{7,\{222\}}=-D_1 D_2 D_3\bigg(p_1p_2p_3\frac{\partial^3}{\partial p_1\partial p_2\partial p_3}\tilde{I}_{1,\{000\}}\bigg),\notag\\
    &\tilde{I}_{5,\{222\}}=D_1D_2D_3 \tilde{I}_{2,\{111\}},\notag\\
    &\tilde{I}_{3,\{222\}}=D_1 D_2 D_3\bigg(\frac{J^2}{4}\tilde{I}_{1,\{000\}}\bigg),\notag\\
    &\tilde{I}_{1,\{222\}}=\big(p_1^2p_2^2p_3^2-\frac{J^2}{4}(p_1^2+p_2^2+p_3^2)\big)\tilde{I}_{1,\{000\}},
\end{align}
where,
\begin{align}
    &D_i=2-p_i\frac{\partial}{\partial p_i},\notag\\
    &\tilde{I}_{2,\{111\}}=-\frac{4p_1^2p_2^2p_3^2}{\mathcal{J}^2}\tilde{I}_{1,\{000\}},\notag\\
    &\tilde{I}_{1,\{000\}}=\frac{1}{\sqrt{-\mathcal{J}^2}}.
\end{align}
If we set $C_1=0$ and $a+c=0$, we get the result corresponding to Einstein gravity in AdS$_5$. On the other hand, setting $a=c=0$ gives the result corresponding to a $\text{Weyl}^3$ interaction. Finally, the coefficient of $a+c$ corresponds to the Gauss-Bonnet contribution.

We also find the following simple double copy relation that determines the Einstein gravity stress tensor correlators given its spin$-1$ Yang-Mills counterpart.
\begin{align}\label{DoublecopyGR}
    \langle T(p_1,\epsilon_1)T(p_2,\epsilon_2)T(p_3,\epsilon_3)\rangle_{\sqrt{-g}R}=(-J^2)^{\frac{1}{2}}\langle J(p_1,\epsilon_1)J(p_2,\epsilon_2)J(p_3,\epsilon_3)\rangle_{YM}^2.
\end{align}
 Finally, the spinor helicity expressions are obtained using \eqref{SHdotprods}.
\section{The Covariant Grassmannian Framework}\label{sec:Covgrassframe}
In this section, we shall construct a Grassmannian framework for correlators in CFT$_4$ involving $\Delta=2$ scalars and conserved currents. Our main result is that these quantities are described by a symplectic bi-Grassmannian structure. This is a doubled Grassmannian with a symplectic orthogonality constraint. We begin by laying out the building blocks of this object and then discuss the explicit construction. We then provide examples of two and three point functions in this framework.

\subsection{The Symplectic Bi-Grassmannian}\label{subsec:covariantsympbigrassman}
We work in the language of off-shell spinor helicity variables. We package all the $\lambda_{i I}^{\alpha}$ and the $\tilde{\lambda}_{iI}^{\Dot{\alpha}}$ into the following vectors:
\begin{align}\label{LambdaandLambdatilde}
    \Lambda_{i I}^{\alpha}\equiv\Lambda=\begin{pmatrix}
        (\lambda_{1})_1^1&(\lambda_{1})_1^2\\
        (\lambda_{1})_2^1&(\lambda_{1})_2^2\\
        \vdots&\vdots\\
        (\lambda_{n})_1^1&(\lambda_{n})_1^2\\
        (\lambda_{n})_2^1&(\lambda_{n})_2^2
    \end{pmatrix},\qquad\tilde{\Lambda}_{iI}^{\Dot{\alpha}}\equiv\tilde{\Lambda}=\begin{pmatrix}
        (\tilde{\lambda}_{1})_1^1&(\tilde{\lambda}_{1})_1^2\\
        (\tilde{\lambda}_{1})_2^1&(\tilde{\lambda}_{1})_2^2\\
        \vdots&\vdots\\
        (\tilde{\lambda}_{n})_1^1&(\tilde{\lambda}_{n})_1^2\\
        (\tilde{\lambda}_{n})_2^1&(\tilde{\lambda}_{n})_2^2
    \end{pmatrix}.
\end{align}
Momentum conservation can then be cast as,
\begin{align}\label{momentumcons}
    \Lambda_{iI}^{\alpha}(\Omega^{ij})^{IJ}\tilde{\Lambda}_{jJ}^{\Dot{\alpha}}\equiv\Lambda^T\cdot\Omega\cdot\tilde{\Lambda}=\sum_{i=1}^{n}\lambda_{i I}^{\alpha}\tilde{\lambda}_{i}^{I\dot\alpha}=0,
\end{align}
where the symplectic form $\Omega$ is given by,
\begin{align}\label{Omega}(\Omega^{ij})^{IJ} = \delta^{ij}\,\epsilon^{IJ} = \begin{pmatrix} \epsilon & 0 & \cdots & 0 \\ 0 & \epsilon & \cdots & 0 \\ \vdots & \vdots & \ddots & \vdots \\ 0 & 0 & \cdots & \epsilon \end{pmatrix}, \quad \epsilon = \begin{pmatrix} 0 & 1 \\ -1 & 0 \end{pmatrix}.\end{align}
The general idea of the Grassmannian framework applied to field theory is to promote the momentum conservation constraint \eqref{momentumcons} to one involving matrices that contain $\Lambda$ and $\tilde{\Lambda}$. In particular, we define the $n\times n\times 2$ matrices $\tilde{C}$ and $C$ that respectively contain $\Lambda$ and $\tilde{\Lambda}$. 
\begin{align}\label{CandCtilde}
C_{ij,I} \equiv C =
\left(
\begin{array}{cc|cc|c|cc}
1_1 & 1_2 & 2_1 & 2_2 & \cdots & n_1 & n_2 \\
\hline
(c_{11})_1 & (c_{11})_2 & (c_{12})_1 & (c_{12})_2 & \cdots & (c_{1n})_1 & (c_{1n})_2 \\
(c_{21})_1 & (c_{21})_2 & (c_{22})_1 & (c_{22})_2 & \cdots & (c_{2n})_1 & (c_{2n})_2 \\
\vdots & \vdots & \vdots & \vdots & \ddots & \vdots & \vdots \\
(c_{n1})_1 & (c_{n1})_2 & (c_{n2})_1 & (c_{n2})_2 & \cdots & (c_{nn})_1 & (c_{nn})_2
\end{array}
\right),
\end{align}
and,
\begin{align}
\tilde C_{ij,I} \equiv \tilde C =
\left(
\begin{array}{cc|cc|c|cc}
\tilde{1}_1 & \tilde{1}_2 & \tilde{2}_1 & \tilde{2}_2 & \cdots & \tilde{n}_1 & \tilde{n}_2 \\
\hline
(\tilde c_{11})_1 & (\tilde c_{11})_2 & (\tilde c_{12})_1 & (\tilde c_{12})_2 & \cdots & (\tilde c_{1n})_1 & (\tilde c_{1n})_2 \\
(\tilde c_{21})_1 & (\tilde c_{21})_2 & (\tilde c_{22})_1 & (\tilde c_{22})_2 & \cdots & (\tilde c_{2n})_1 & (\tilde c_{2n})_2 \\
\vdots & \vdots & \vdots & \vdots & \ddots & \vdots & \vdots \\
(\tilde c_{n1})_1 & (\tilde c_{n1})_2 & (\tilde c_{n2})_1 & (\tilde c_{n2})_2 & \cdots & (\tilde c_{nn})_1 & (\tilde c_{nn})_2
\end{array}
\right).
\end{align}
We further impose the constraint,
\begin{align}\label{CdotCtilde}
    \tilde{C}_{ij,I}(\Omega^{jk})^{IJ}C_{kl,J}=\tilde{C}^T\cdot\Omega\cdot C=0,
\end{align}
along with the conditions,
\begin{align}\label{CdotLambda}
   &\tilde{C}_{ij,I}(\Omega^{jk})^{IJ}\Lambda^{\alpha}_{kJ}= \tilde{C}^T\cdot \Omega\cdot \Lambda=0,\notag\\
   &C_{ij,I}(\Omega^{jk})^{IJ}\tilde{\Lambda}^{\Dot{\alpha}}_{kJ}= C^T\cdot \Omega\cdot \tilde{\Lambda}=0.
\end{align}
It is easy to show that \eqref{CdotCtilde} and \eqref{CdotLambda} imply momentum conservation \eqref{momentumcons}. With the above at hand, we are ready to define the Grassmannian.

\begin{center}
\fbox{%
\begin{minipage}{0.97\textwidth}
\textbf{The Symplectic Bi-Grassmannian}
We now combine the kinematic constraints into an integral representation.

\paragraph{Definition.}
We define the symplectic bi-Grassmannian representation of the $n$-point correlator as
\begin{align}\label{GrassmannianSH}
\Psi_n^{\{I_1,\dots\},\dots,\{I_n,\dots\}}
=&
\int \frac{dC}{\mathrm{Vol}(\mathrm{GL}(n))}
\int\frac{d\tilde C}{\mathrm{Vol}(\mathrm{GL}(n))}
\;\nonumber\\
&\times 
\delta(\tilde C^T \Omega C)\,
\delta(\tilde C^T \Omega \Lambda)\,
\delta(C^T \Omega \tilde\Lambda)\,
A_n^{\{I_1,\dots\},\dots,\{I_n,\dots\}}(C,\tilde C).
\end{align}
{This should be contrasted with the symplectic Grassmannian integral appearing in Coulomb-branch amplitudes \cite{CalvoCortes:2025Suthar} in 4-d, where one integrates over a single $n$-plane $C\in SpGr(n,2n)$ satisfying $C\Omega C^T=0$ as a result of special kinematics. In the present CFT$_4$ construction, the basic object is instead a pair of $n$-planes $(C,\widetilde C)$ with the mutually orthogonal constraint $\widetilde C^T\Omega C=0$.}

Let us note that
\begin{itemize}
\item $C$ and $\tilde C$ are $n \times 2n $ matrices with a 
$\mathrm{GL}(n)\times\mathrm{GL}(n)$ redundancy.
\item They are constrained to be symplectically orthogonal:
\[
\tilde C^T \Omega C = 0.
\]
\item The function $A_n(C,\tilde C)$ depends on the $n \times n$ minors of $C$ and $\tilde C$, 
which carry $\mathrm{GL}(2,\mathbb{R})$ indices.
\end{itemize}

\end{minipage}%
}
\end{center}
\begin{center}
\fbox{%
\begin{minipage}{0.97\textwidth}
\paragraph{GL$(n)\times$GL$(n)$ invariance.}
Under
\[
C \to G\,C, \qquad \tilde C \to \tilde G\,\tilde C,
\]
the invariance of the integrand requires
\begin{equation}\label{Ascal}
A_n(C,\tilde C)\to
\det(G)^{2-n}\det(\tilde G)^{2-n}\,A_n(C,\tilde C).
\end{equation}

The object $\Psi_n$ describes $n$-point correlators of either $\Delta=2$ scalar operators 
or symmetric traceless spin-$s$ conserved currents. The little group covariance of the 
external operators constrains the allowed dependence of $A_n(C,\tilde C)$ on the minors.
A key feature of this construction is that, the delta-function constraints ensure invariance under the full conformal group.
Geometrically, the matrices $C$ and $\tilde C$ define two $n$-planes in a 
$2n$-dimensional symplectic vector space, constrained to be mutually isotropic and 
aligned with the external kinematic data.
\end{minipage}%
}
\end{center}

Let us now discuss the construction the $\mathbb{GL}(n)$ invariant minors of the $C$ and $\tilde{C}$ matrices. From \eqref{CandCtilde}, we see that the first two columns of $C$ are labeled by $1_I$, the second two by $2_I$ and so on. Similarly for $\tilde{C}$, the first two columns are labeled by $\tilde{1}_I$, the second two by $\tilde{2}_I$ etc.. This is because these are representative of the way they transform under the $\mathbb{GL}(2)_1\times\mathbb{GL}(2)_2\cdots$ little-group transformations of the external kinematic data. This allows us to construct  $A_n^{\{I_1,\cdots\},\cdots,\{I_n,\cdots\}}(C,\tilde{C})$ in a manifestly $GL(2,\mathbb{R})$ covariant manner by constructing it out of minors that themselves carry $SL(2,\mathbb{R})$ indices that ensure correct transformation properties for each operator viz \eqref{littlegroupscaling}. Further, we must arrange for columns without tildes and with tildes in such a way to ensure the correct $GL(1)$ properties of each operator \eqref{littlegroupscaling}. Under a little group transformation \eqref{littlegroupscaling} we have,
\begin{align}\label{ianditildetransforms}
    &(\cdots, i^{I},\cdots)\to \frac{1}{r_i^{m_i}}(S_i)^{I}_{J}(\cdots,i^{J},\cdots),\notag\\
    &(\cdots, \tilde{i}^{I},\cdots)\to \frac{1}{r_i^{n_i}}(S_i)^{I}_{J}(\cdots,i^{J},\cdots),
\end{align}
and where $m_i+n_i=s_i$ as we constructed in \eqref{Jsrescaled}.
\subsection{Proof of conformal invariance}
We now check the conformal invariance of our construction \eqref{GrassmannianSH}. We see that the delta functions impose,
\begin{align}
    \tilde{C}^T\cdot\Omega\cdot C=0,~C^T\cdot\Omega\cdot\tilde{\Lambda}=0,~\tilde{C}^T\cdot\Omega\cdot\Lambda=0.
\end{align}
The second and third conditions imply that $\tilde{\Lambda}$ and $\Lambda$ are symplectically orthogonal to $C$ and $\tilde{C}$ respectively. Given the fact that $C$ and $\tilde{C}$ are also symplectically orthogonal, it is easy to see that $\tilde{\Lambda}^T\cdot\Omega\cdot\Lambda=0$ which is the statement of momentum conservation. 

As for invariance under special conformal transformations we note that the SCT operator can be written as,
\begin{align}
    K_{\alpha\Dot{\alpha}}=\frac{\partial}{\partial \Lambda_{i}^{I\alpha}}(\Omega^{ij})^{IJ}\frac{\partial}{\partial\tilde{\Lambda}_{j}^{J\Dot{\alpha}}}.
\end{align}
The derivatives act only on the factors $\delta(\tilde{C}\cdot\Omega\cdot \Lambda)\delta(C\cdot\Omega\cdot\tilde{\Lambda})$ in the Grassmannian \eqref{GrassmannianSH}. This results in,
\begin{align}
    \tilde{C}^T\cdot \Omega\cdot\Omega\cdot\Omega\cdot C\propto \tilde{C}^T\cdot\Omega\cdot C=0,
\end{align}
by the symplectic orthogonality constraint. We also used the fact that $\Omega$ squares to the negative identity matrix.
Now, we proceed to bootstrap $A(C,\tilde{C})$ for various examples which upon performing the Grassmannian integrals, reproduce the spinor helicity results of the previous section.

\subsection{Two point functions}\label{sec:2pointcovariantgrass}
Two point function is given by putting n=2 in \eqref{GrassmannianSH} 
\footnotesize
\begin{align}\label{GrassmannianSH2pt}
    \Psi_2^{\{I_1\cdots\},\cdots\{I_n,\cdots\}}=\int \frac{d^{2\times 2\times 2} C}{\text{Vol}(GL(2))}\int\frac{d^{2\times 2 \times 2} \tilde{C}}{\text{Vol}(GL(2))}\delta^{2\times 2}(\tilde{C}^T\cdot \Omega\cdot C)\delta(\tilde{C}^T\cdot\Omega\cdot \Lambda)\delta(C^T\cdot\Omega\cdot\tilde{\Lambda})A_2^{\{I_1,\cdots\}\cdots\{I_n,\cdots\}}(C,\tilde{C}).
\end{align}
where 
\normalsize
$A_2^{\{I_1,\cdots\}\cdots\{I_n,\cdots\}}(C,\tilde{C})$ satisfies,
\begin{align}
    A_2^{\{I_1,\cdots\}\cdots\{I_n,\cdots\}}(GC,\tilde{G}\tilde{C})=A_2^{\{I_1,\cdots\}\cdots\{I_n,\cdots\}}(C,\tilde{C}),~(G,G')\in GL(2)\times GL(2)
\end{align}, see \eqref{Ascal}.
We must also construct this function using the properties \eqref{ianditildetransforms}, such that the $GL(2)$ redundancy of each external operator \eqref{littlegroupscaling} is respected. Before we proceed to the examples, let us discuss the gauge-fixing procedure for two points. First, we can solve for $\tilde{C}$ in terms of the components of $C$ using the symplectic orthogonality constraint. We gauge-fix,
\begin{align}
    C=\begin{pmatrix}
        1&0&c_{12}&-c_{11}\\
        0&1&c_{22}&-c_{21}\end{pmatrix},~\tilde{C}=\begin{pmatrix}
            a_{11}&a_{12}&-1&0\\
            a_{21}&a_{22}&0&-1
        .
    \end{pmatrix}.
\end{align}
$\tilde{C}^T\cdot \Omega \cdot C=0$ then fixes,
\begin{align}
    \tilde{C}=\begin{pmatrix}
       -c_{21}&c_{11}&-1&0\\
        -c_{22}&c_{12}&0&-1
    \end{pmatrix}.
\end{align}
The resulting gauge-fixed Grassmannian is,
\begin{align}\label{2ptGrassmanniangaugefixed}
    \Psi^{\{I_1,\cdots,\},\cdots,\{I_n,\cdots,\}}=\int d^4 c_{IJ}\delta^4(\tilde{\lambda}_1^{I \Dot{\alpha}}+c_{IJ}\tilde{\lambda}_{2J}^{\Dot{\alpha}})\delta^4(\lambda_2^{I\alpha}+c_{JI}\lambda_{1}^{J\alpha})A^{\{I_1,\cdots\},\cdots,\{I_n,\cdots\}}(C,\tilde{C}),
\end{align}
where $C$ and $\tilde{C}$ now denote the gauge-fixed matrices.
\subsubsection{$\langle O_2O_2\rangle$}
We begin with the simplest two point function of a $\Delta=2$ scalar operator. In momentum space, it is a constant \eqref{O2O2SH}. In the Grassmannian \eqref{GrassmannianSH2pt}, it turns out it is also a constant. For two point functions, $A(C,\tilde{C})$ should be invariant with respect to a $GL(2)\times GL(2)$ transformation of the matrices $C$ and $\tilde{C}$. Since the external operators are scalars, we must also have invariance under individual little group transformations $GL(2)_1\times GL(2)_2$ for each operator. The simplest function that satisfies this is,
\begin{align}
    A_2(C,\tilde{C})=1.
\end{align}
One can check using \eqref{2ptGrassmanniangaugefixed} that this results in the correct momentum space answer viz \eqref{O2O2SH}.

\subsubsection{$\langle O_{\frac{1}{2}}^{I}\tilde{O}_{\frac{1}{2}}^{J}\rangle$}
For the spin$-\frac{1}{2}$ case, we must now take into account the non-trivial transformation of each operator under the $GL(1)_1\times GL(1)_2$ subgroup of the little group \eqref{littlegroupscaling}. In particular we have $m=1,n=0$ for $O_{\frac{1}{2}}$ and $m=0,n=1$ for $\tilde{O}_{\frac{1}{2}}$ as we can see from the definition \eqref{spinhalfI}. Further, the ansatz we write down has to be invariant under $GL(2)\times GL(2)$ transformations corresponding to the $C$ and $\tilde{C}$ matrices. Thus, the answer has to be formed out of a ratio of minors. A natural ansatz is,
\begin{align}
    A_2^{I,J}(C,\tilde{C})=\frac{(1^{I}1_K)(\tilde{1}^K \tilde{2}^J)}{(1_K 2_J)(\tilde{1}^K \tilde{2}^J)}.
\end{align}
Under a little group transformation, the minors transform as in \eqref{ianditildetransforms}. The free $I$ and $J$ indices take care of the $SL(2)_1\times SL(2)_2$ covariance of the external operators. As for the external $GL(1)_1\times GL(1)_2$ covariance, we find using \eqref{ianditildetransforms} that we get an overall scaling of $\frac{r_2}{r_1}$ which is perfectly consistent with \eqref{littlegroupscaling}. 

Finally, we can gauge-fix as in \eqref{2ptGrassmanniangaugefixed}. We find that the numerator becomes $c^{IJ}$ while the denominator becomes $\text{Det}(c)$. Thus we get,
\begin{align}
    \Psi^{I,J}=\int d^4 c_{IJ}\delta^4(\tilde{\lambda}_1^{I \Dot{\alpha}}+c_{IJ}\tilde{\lambda}_{2J}^{\Dot{\alpha}})\delta^4(\lambda_2^{I\alpha}+c_{JI}\lambda_{1}^{J\alpha})\frac{c^{IJ}}{\text{Det}(c)}.
\end{align}
Performing the integral results in the correct result \eqref{OhOhSH}.

\subsubsection{$\langle J^{I_1J_1}J^{I_2J_2}\rangle$}
We find the following expression for the spin$-1$ two point function.
\begin{align}
    A_2^{(I_1,J_1),(I_2,J_2)}(C,\tilde{C})=\frac{(1^{(I_1}2^{(I_2})(\tilde{1}^{J_1)}\tilde{2}^{J_2)})}{(1_I 2_J)(\tilde{1}^I\tilde{2}^J)}.
\end{align}
Gauge-fixing \eqref{2ptGrassmanniangaugefixed} we find,
\begin{align}
    \Psi_2^{(I_1 J_1)\cdots(I_2 J_2)}=\int d^4 c_{IJ}\delta^4(\tilde{\lambda}_1^{I \Dot{\alpha}}+c_{IJ}\tilde{\lambda}_{2J}^{\Dot{\alpha}})\delta^4(\lambda_2^{I\alpha}+c_{JI}\lambda_{1}^{J\alpha})\frac{c^{(I_1 (I_2}c^{J_1) J_2)}}{\text{Det}(c)}.
\end{align}
Performing the integral and converting to momentum space results in the correct expression \eqref{JJSH}. Proceeding this way, one can easily generalize the results to arbitrary spin.

\subsection{Three point functions}\label{sec:3pointcovariantgrass}
We now move on to the more involved case of three point functions. We have for $n=3$ \eqref{GrassmannianSH},
\begin{align}
    \Psi_3^{(I_1\cdots),\cdots,(I_n,\cdots)}=\int \frac{d^{3\times 3 \times 2} C}{\text{Vol}(GL(3))}\int \frac{d^{3\times 3 \times 2} \tilde{C}}{\text{Vol}(GL(3))}\delta(\tilde{C}^T\cdot \Omega\cdot C)\delta(\tilde{C}\cdot\Lambda)\delta(C\cdot \tilde{\Lambda})A_3^{(I_1\cdots),\cdots,(I_n,\cdots)}
\end{align}
At the level of three points, we require,
\begin{align}\label{3ptGL3Gl3}
    A_3(GC,\tilde{G}\tilde{C})=\frac{1}{\text{Det}(G)\text{Det}(\tilde{G})}A_3(C,\tilde{C}),
\end{align}
to ensure the $GL(3)\times GL(3)$ invariance of the symplectic Grassmannian.
To determine $A_3$, we first write down an ansatz consistent with \eqref{3ptGL3Gl3}. We then demand the correct $GL(2,\mathbb{R})$ covariance  with respect to each external operator \eqref{littlegroupscaling}. This involves suitably choosing the minor tensors with correct index placement and such that the minors of $C$ and $\tilde{C}$ appear in such a way to ensure the external $GL(1)\times GL(1)\times GL(1)$ invariance. There may be many seemingly in-equivalent ansatz satisfying these properties but under the support of $\tilde{C}^T\cdot \Omega\cdot C$, many of these are equivalent and thus we obtain the correct number of independent structures for each correlator we study. Finally, we have verified that the expressions presented below after performing the Grassmannian integrals, result in the correct spinor helicity variables results given in the previous section. Before we proceed, we define the following unique little group invariant quantity at three points that will be useful in what follows,
\begin{align}\label{threepointK}
    \frac{1}{\mathcal{K}}=\frac{1}{(1^I 1^J 2^K)(\tilde{1}_I\tilde{1}_J\tilde{2}_K)}=\frac{1}{(2^I 2^J 3^K)(\tilde{2}_I\tilde{2}_J\tilde{3}_K)}=\frac{1}{(3^I 3^J 1^K)(\tilde{3}_I\tilde{3}_J\tilde{1}_K)},
\end{align}
where all the ansatz are equal at the support of the symplectic constraint ($\tilde C^T\cdot \Omega\cdot {C} = 0$) on the Grassmannian.

\subsubsection{$\langle O_2O_2O_2\rangle$}
Our ansatz must satisfy \eqref{3ptGL3Gl3}. Since the external operators are scalars, we also need the ansatz to be invariant under the external $GL(2)$ little group transformations for each operator. This heavily restricts its allowed form. We find that the following expression is sufficient to reproduce the correct momentum space result.
\begin{align}
    A_3(C,\tilde{C})=\frac{Sgn(\mathcal{K})}{\mathcal{K}},
\end{align}
where $\mathcal{K}$ is the unique little group invariant \eqref{threepointK}\footnote{A naive scalar ansatz \(1/\Block \) has the wrong Bose symmetry: it is odd under exchange of two scalar insertions. Since the momentum-space scalar Wightman function is symmetric, the scalar Grassmannian integrand must be made symmetric too. We therefore use $A^{OOO}_3(C,\tilde C) = \frac{Sgn(\Block)}{\Block} = \frac1{|K|} $ on the real Grassmannian slice. However, while doing the Grassmannian integral for three point we will have to do one integral and to do that we choose a $i \epsilon$ prescription by making $\Block \xrightarrow{} \Block+ i \epsilon$ by choosing to close the contour in the upper half plane. Please note $Sgn(\Block+ i \epsilon)\sim 1$ for $\epsilon \xrightarrow{}0$. Please refer to last part of this section to see the discussion elaborately.}.
Any other ansatz is related to this one by Plucker relations or the symplectic orthogonality constraint. We have also checked that this ansatz leads to the correct momentum space result.

\subsubsection{$\langle J^{IJ}O_2O_2\rangle$}
We now look for an ansatz that has two free indices. A natural choice also satisfying the $GL(3)\times GL(3)$ covariance is,
\begin{align}
    A_3^{\{IJ\}}=\frac{(1^{(I} 2^K 2^L)(\tilde{1}^{J)} \tilde{2}_K\tilde{2}_L)}{\mathcal{K}^2}\text{sgn}(\mathcal{K}).
\end{align}
The $I,J$ indices are symmetrized since the are indices of the spin$-1$ conserved current $J^{IJ}$. The indices of the columns corresponding to $2$ and $\tilde{2}$ are contracted to ensure $SL(2)$ invariance of the second operator. Finally, although the ansatz does not have manifest $2\leftrightarrow 3$ exchange (anti-)symmetry, it is a simple matter to check that on the support of the delta functions in the Grassmannian integral, it is indeed there. The factor of $\text{sgn}(\mathcal{K})$ is essential to ensure this permutation (anti-)symmetry.

\subsubsection{$\langle T^{(IJKL)}O_2O_2\rangle$}
We consider a similar correlator now with a stress tensor instead of a spin$-1$ current and two scalars. A natural ansatz based on the previous $\langle JO_2O_2\rangle$ example is the following:
\begin{align}
    A_3^{\{I_1J_1K_1L_1\}}=\frac{(1^{I_1}2^I2^K)(\tilde{1}^{J_1}\tilde{2}_I\tilde{2}_K)(1^{K_1}2^I 2^K)(\tilde{1}^{L_1}\tilde{2}_I\tilde{2}_K)}{\mathcal{K}^3}\text{Sgn}(\mathcal{K}),
\end{align}
with an implicit symmetrization in the $(I_1,J_1,K_1,L_1)$ indices. In contrast to the spinor helicity space expressions \eqref{JOOSH} and \eqref{TOOSH}, the Grassmannian result for $\langle TO_2O_2\rangle$ is simply a product of two copies of $\langle JO_2O_2\rangle$ multiplied by $|\mathcal{K}|$ to obtain the correct $GL(3)\times GL(3)$ covariance.
\subsubsection{$\langle O_{\frac{1}{2}}^I\tilde{O}_{\frac{1}{2}}^JO_2\rangle$}
We now move on to examples involving two operators with non-zero spin. We find,
\begin{align}
    A_3^{\{I\},\{J\}}=\frac{(\tilde{1}^{I}\tilde{2}^J\tilde{3}^K)(2_M 2^M 3_K)\text{Sgn}(\mathcal{K})}{\mathcal{K}^2}.
\end{align}
Note that this quantity under a $GL(1)$ little group transformation of the external operators has a non-trivial weight which is in accordance to what we expect of spin$-\frac{1}{2}$ operators. We note the simplicity of this result over its spinor helicity counterpart \eqref{OhOhO2SH}.
\subsubsection{$\langle J^{I_1J_1}J^{I_2J_2}O_2\rangle$}
This correlator can be obtained by a double-copy like structure from $\langle O_{\frac{1}{2}}\tilde{O}_{\frac{1}{2}}O_2\rangle$. However, we need to take into account the fact that the former correlator has non-trivial $GL(1)$ little group scaling with respect to the spin$-\frac{1}{2}$ operators. Thus, we construct the following combination to ensure the $GL(1)$ little group invariances of the spin$-1$ currents.
\begin{align}
    A_3^{\{I_1J_1\},\{I_2J_2\}}=Sgn(\mathcal{K})\mathcal{K}\frac{(\tilde{1}^{I_1}\tilde{2}^{I_2}\tilde{3}^K)(2_M 2^M 3_K)}{\mathcal{K}^2}\frac{(1^{J_1}2^{J_2}3^K)(\tilde{2}_M \tilde{2}^M \tilde{3}_K)}{\mathcal{K}^2}.
\end{align}
One can verify that this result goes over the correct spinor helicity expression \eqref{JJO2SH} upon performing the Grassmannian integrals.
\subsubsection{$\langle J^{I_1J_1}J^{I_2J_2}J^{I_3J_3}\rangle$}
We now move on to the interesting example of the correlator of three spin$-1$ currents. As we discussed in the previous section, there are two independent structures consistent with conformal invariance which correspond to Yang-Mills theory \eqref{JJJYM} and a $F^3$ contribution \eqref{JJJF3} from the AdS$_5$ bulk perspective. In the Grassmannian, we find the following expressions for the correlators corresponding to these interactions.
\begin{align}\label{YMGrassmannian}
    A_{3,YM}^{\{I_1J_1\},\{I_2J_2\},\{I_3J_3\}}=\frac{(1^{I_1}2^{I_2}3^{I_3})(\tilde{1}^{J_1}\tilde{2}^{J_2}\tilde{3}^{J_3})}{\mathcal{K}^2}\text{Sgn}(\mathcal{K}).
\end{align}
The factor of $\text{Sgn}(\mathcal{K})$ ensures anti-symmetry with respect to $(1,2,3)$. The symmetrization of the indices of each current is implicit. This is compensated by the anti-symmetry of the structure constants $f^{ABC}$ of the non-abelian group to ensure Bose-symmetry of the correlator as a whole. Yet again, compared to its spinor helicity counterpart \eqref{JJJYM}, this expression is much more simple consisting of just one term rather than three.

Next, we move on to the $F^3$ contribution. Here, we find an interesting structure.
\footnotesize
\begin{align}\label{F3Grassmannian}
     &A_{F^3}^{\{I_1J_1\},\{I_2J_2\},\{I_3J_3\}}\notag\\&=\bigg(\frac{3(1^{I_1} 2_K 2^K)(\tilde{1}^J_1 \tilde{2}_K \tilde{2}^K)(2^{I_2}3_K3^K)(\tilde{2}^{J_2}\tilde{3}_K\tilde{3}^K)(3^{I_3}1_K1^K)(\tilde{3}^{J_3}\tilde{1}_K\tilde{1}_K)}{32 \mathcal{K}^4}-\frac{(1^{I_1}2^{I_2}3^{I_3})(\tilde{1}^{J_1}\tilde{2}^{J_2}\tilde{3}^{J_3})}{\mathcal{K}^2}\bigg)\text{Sgn}(\mathcal{K})\notag\\
     &=\frac{3(1^{I_1} 2_K 2^K)(\tilde{1}^J_1 \tilde{2}_K \tilde{2}^K)(2^{I_2}3_K3^K)(\tilde{2}^{J_2}\tilde{3}_K\tilde{3}^K)(3^{I_3}1_K1^K)(\tilde{3}^{J_3}\tilde{1}_K\tilde{1}_K)}{32 \mathcal{K}^4}\text{Sgn}(\mathcal{K})-A_{3,YM}^{\{I_1J_1\},\{I_2J_2\},\{I_3J_3\}}.
\end{align}
\normalsize
The first term in the above expression is the sum of the $YM$ and $F^3$ contributions and thus we remove the $YM$ contribution which we independently computed \eqref{YMGrassmannian} to obtain the $F^3$ result \eqref{F3Grassmannian}.

\subsubsection{$\langle T^{I_1J_1K_1L_1}J^{I_2J_2}J^{I_3J_3}\rangle$}
We now consider a correlator of a stress tensor and two spin$-1$ currents. We find the following general ansatz in the Grassmannian space.
\begin{align}
    &A_3^{\{I_1J_1K_1L_1\},\{I_2J_2\},\{I_3J_3\}}\notag\\&=\alpha\Bigg[\frac{\text{Sgn}(\mathcal{K})}{2^{6}\mathcal{K}^5}(1^{I_1}2_K2^K)(\tilde{1}^{J_1}\tilde{2}^K\tilde{2}_K)(2^{I_2}3_K3^K)(\tilde{2}^{J_2}\tilde{3}_K\tilde{3}^K)(3^{I_3}1_K1^K)(\tilde{3}^{K_3}\tilde{1}_K\tilde{1}^K)(1^{K_1}2_K2^K)(\tilde{1}^{L_1}\tilde{2}_K\tilde{2}^K)\Bigg]\notag\\
    &+\beta\Bigg[\frac{\text{Sgn}(\mathcal{K})}{\mathcal{K}^3}(1^{I_1}2^{I_2}3^{I_3})(\tilde{1}^{J_1}\tilde{2}^{J_2}\tilde{3}^{J_3})(1^{K_1}2_K2^K)(\tilde{1}^{L_1}\tilde{2}_K\tilde{2}^K)\Bigg].
\end{align}
The symmetrization of the indices of the stress tensor and the currents are implicit.
Converting to spinor helicity and matching with the result \eqref{TJJSH} yields the values,
\begin{align}
    \alpha=-12C_1,\beta=C_J-2C_1.
\end{align}

\subsubsection{$\langle T^{I_1J_1K_1L_1}T^{I_2J_2K_2L_2}T^{I_3J_3K_3L_3}\rangle$}
Finally, we consider the correlator of three stress tensors. As we discussed, this correlator has three linearly independent tensor structures characterized by the coefficients $C_1,c$ and $a$. To construct these quantities in the Grassmannian, we take motivation from $\langle JJJ\rangle$. We have \eqref{YMGrassmannian} and the first term in \eqref{F3Grassmannian} as independent structures for the spin$-1$ three point function. Taking a product of two copies of \eqref{YMGrassmannian} (indices appropriately inserted to be consistent with symmetrization) and multiplying by $|\mathcal{K}|$ produces a consistent spin-$2-2-2$ object. We can also take a product of two copies of the first term in \eqref{F3Grassmannian} or even one copy of \eqref{YMGrassmannian} and \eqref{F3Grassmannian}. Thus we have the general ansatz,
\begin{align}\label{TTTGrassmannian}
    &A_3^{\{I_1J_1K_1 L_1\},\{I_2J_2K_2 L_2\},\{I_3J_3K_3 L_3\}}\notag\\&=\alpha\Bigg[\frac{\text{Sgn}(\mathcal{K})}{2^{12}\mathcal{K}^7}(1^{I_1}2_K2^K)(\tilde{1}^{J_1}\tilde{2}_K\tilde{2}^K)(1^{K_1}2_K2^K)(\tilde{1}^{L_1}\tilde{2}_K\tilde{2}^K)(2^{I_2}3_K3^K)(\tilde{2}^{J_2}\tilde{3}_K\tilde{3}^K)(2^{K_2}3_K3^K)(\tilde{2}^{L_2}\tilde{3}_K\tilde{3}^K)\notag\\&\times(3^{I_3}1_K1^K)(\tilde{3}^{J_3}\tilde{1}_K\tilde{1}^K)(3^{K_3}1_K1^K)(\tilde{3}^{L_3}\tilde{1}_K\tilde{1}^K)\Bigg]\notag\\
    &+\beta\Bigg[\frac{\text{Sgn}(\mathcal{K})}{\mathcal{K}^3}(1^{I_1}2^{I_2}3^{I_3})(\tilde{1}^{J_1}\tilde{2}^{J_2}\tilde{3}^{J_3})(1^{K_1}2^{K_2}3^{K_3})(\tilde{1}^{L_1}\tilde{2}^{L_2}\tilde{3}^{L_3})\Bigg]\notag\\
    &+\gamma\Bigg[\frac{\text{Sgn}(\mathcal{K})}{2^{6}\mathcal{K}^5}(1^{I_1}2_K2^K)(\tilde{1}^{J_1}\tilde{2}_K\tilde{2}^K)(2^{I_2}3_K3^K)(\tilde{2}^{J_2}\tilde{3}_K\tilde{3}^K)(3^{I_3}1_K1^K)(\tilde{3}^{J_3}\tilde{1}_K\tilde{1}^K)(1^{K_1}2^{K_2}3^{K_3})(\tilde{1}^{L_1}\tilde{2}^{L_2}\tilde{3}^{L_3})\Bigg].
\end{align}
The symmetrization of the indices of the stress tensors is implicit. 
Converting the above ansatz into spinor helicity variables and comparing with the result \eqref{TTTSH}, we find,
\begin{align}
    &\alpha=-720 C_1,\notag\\
    &\beta=-8 C_1+2c+4a,\notag\\
    &\gamma=-192 C_1+24(a+c).
\end{align}
For example, consider Einstein gravity. We have $a+c=0$ and $C_1=0$ which shows only the coefficient of $\beta$ contributes which is the Yang-Mills ``squared", consistent with the double copy in momentum space \eqref{DoublecopyGR}. Let us now take a moment to compare the momentum space expression \eqref{TTTSH} with the Grassmannian result \eqref{TTTGrassmannian}. The latter is extremely simple compared to the former, thus highlighting the utility of the Grassmannian framework.

\subsection{Back to spinor helicity}
To verify our gauge-covariant results, we will take the following gauge choice. 
\begin{align}
    C=\begin{pmatrix}
        1&0&0&c_{11}&c_{12}&c_{13}\\
        0&1&0&c_{21}&c_{22}&c_{23}\\
        0&0&1&c_{31}&c_{32}&c_{33},
    \end{pmatrix}~~\tilde{C}=\begin{pmatrix}
        a_{11}&a_{12}&a_{13}&-1&0&0\\
        a_{21}&a_{22}&a_{23}&0&-1&0\\
        a_{31}&a_{32}&a_{33}&0&0&-1
    \end{pmatrix}.
\end{align}
The symplectic orthogonality condition then implies that,
\begin{align}
    \tilde{C}=\frac{1}{c_{31}}\begin{pmatrix}
        c_{21}&-c_{11}&-1&-c_{31}&0&0\\
        c_{23}c_{31}-c_{21}c_{33}&-c_{13}c_{31}+c_{11}c_{33}&c_{33}&0&-c_{31}&0\\
        -c_{22}c_{31}+c_{21}c_{32}&c_{12}c_{31}-c_{11}c_{32}&-c_{32}&0&0&-c_{31}
    \end{pmatrix}.
\end{align}
This also produces a Jacobian factor of $\frac{1}{c_{31}^3}$. To evaluate the three point Grassmannian integrals, it is useful to gauge-fix the $GL(2)$ redundancies of the external operators as well. This does not lose any generality as we can easily uplift this to obtain the expressions in a general gauge easily. We make the convenient choice,
\begin{align}
     &\tilde{\lambda}_{1}^{I\Dot{\alpha}}=\delta^{I\Dot{\alpha}},\notag\\
    &\lambda_{2}^{I\alpha}=\delta^{I\alpha},\notag\\
    &\lambda_3^{I\alpha}=\delta^{I\alpha},
\end{align}
and,
\begin{align}
    &\lambda_{1}^{1\alpha}=\lambda_1^\alpha,\lambda_{1}^{2\alpha}=\rho_1^\alpha,\notag\\
    &\tilde{\lambda}_2^{1\Dot{\alpha}}=\tilde{\lambda}_2^{\Dot{\alpha}},\tilde{\lambda}_2^{2\Dot{\alpha}}=\tilde{\rho}_2^{\Dot{\alpha}}\notag\\
    &\tilde{\lambda}_3^{1\Dot{\alpha}}=\tilde{\lambda}_3^{\Dot{\alpha}},\tilde{\lambda}_3^{2\Dot{\alpha}}=\tilde{\rho}_3^{\Dot{\alpha}}.
\end{align}
We then find,
\begin{align}
    \delta(C\cdot\tilde{\Lambda})=&\delta^2(\delta^{\Dot{\alpha}1}+c_{12}\tilde{\lambda}_3^{\Dot{\alpha}}+c_{11}\tilde{\rho}_2^{\Dot{\alpha}}+c_{13}\tilde{\rho}_3^{\Dot{\alpha}})\notag\\
    &\delta^2(\delta^{\Dot{\alpha}2}+c_{22}\tilde{\lambda}_3^{\Dot{\alpha}}+c_{21}\tilde{\rho}_2^{\Dot{\alpha}}+c_{23}\tilde{\rho}_3^{\Dot{\alpha}})\notag\\
    &\delta^2(\tilde{\lambda}_2^{\Dot{\alpha}}+c_{32}\tilde{\lambda}_3^{\Dot{\alpha}}+c_{31}\tilde{\rho}_2^{\Dot{\alpha}}+c_{33}\tilde{\rho}_3^{\Dot{\alpha}}),
\end{align}
and,
\begin{align}
    \delta(\tilde{C}\cdot\Lambda)=&\delta^2(c_{21}\lambda_1^{\alpha}-c_{11}\rho_1^{\alpha}-\frac{\delta^{\alpha 1}}{c_{31}}-\delta^{\alpha 2})\notag\\
    &\delta^2((c_{23}-\frac{c_{21}c_{33}}{c_{31}})\lambda_1^\alpha+(-c_{13}+\frac{c_{11}c_{33}}{c_{31}})\rho_1^\alpha+(\frac{c_{33}}{c_{31}}-1)\delta^{\alpha 1})\notag\\
    &\delta^2((-c_{22}+\frac{c_{21}c_{32}}{c_{31}}\lambda_1^\alpha+(c_{12}-\frac{c_{11}c_{32}}{c_{31}})\rho_1^\alpha-\frac{c_{32}\delta^{\alpha 1}}{c_{31}}-\delta^{\alpha 2}).
\end{align}
Thus we find,
\begin{align}
    &\Psi_3^{(I_1\cdots),(I_n,\cdots)}\notag\\&=\frac{1}{|\lambda_1\cdot\rho_1|~|\tilde{\lambda}_3\cdot\tilde{\rho}_3|}\int dc_{11}dc_{12}dc_{13}dc_{21}dc_{22}dc_{23}dc_{31}dc_{32}dc_{33}\frac{1}{|c_{31}|}\delta(C\cdot\tilde{\Lambda})\delta(\tilde{C}\cdot\Lambda)A^{(I_1\cdots),(I_n,\cdots)},
\end{align}
where the delta functions are given as above and $C$ and $\tilde{C}$ are the gauge-fixed matrices. We have $12$ delta functions and $9$ integrals. $4$ of the $12$ delta functions conspire to turn into the momentum conserving $\delta^4(p_1+p_2+p_3)$. The remaining $8$ can be used to evaluate all but one integral, say over $c_{31}$. We perform this last integral by complexifying and converting it into a contour integral. We remove the absolute values and use the $i\epsilon$ prescription $c_{31}\to c_{31}+i\epsilon$ to evaluate the integral which generically has two poles. This prescription chooses one of the two poles and using it, we have verified that all the Grassmannian expressions given in this section reproduce the correct spinor helicity results.

\section{The Helicity basis Grassmannian Framework}\label{sec:HBgrassframework}
Prior to the above construction, where operators were covariantly the representation of the little group, here we discuss the helicity basis in the component language and the operators as their representation.  

We start with the spinor helicity variables for 4-momentum in the component language using eq\eqref{PinSH}, 
\begin{align}\label{eq:momentumSH}
p_{\mu}\sigma^\mu_{\alpha\dot\alpha}=p_{\alpha\dot\alpha}=\lambda_\alpha^I\epsilon_{IJ}\tilde\lambda^J_{\dot\alpha}= \lambda_\alpha \tilde\rho_{\dot\alpha}-\rho_\alpha \tilde\lambda_{\dot\alpha}
\end{align}
where, $\lambda_\alpha^I=\{\lambda_\alpha,\rho_\alpha\}$ and $\tilde\lambda_{\dot\alpha}^I=\{\tilde\rho_{\dot\alpha}, \tilde\lambda_{\dot\alpha}\}$ are in the fundamental of $SL(2,\mathbb{R})_L$ and $SL(2,\mathbb{R})_R$ respectively, with the $\alpha$ and $\dot\alpha$ index respectively. The little group for the above given momentum, in spinor representation, is $SL(2,\mathbb R) \times GL(1,\mathbb{R})$ whose infinitesimal differential and matrix representation is disscussed in detail in the Appendix\ref{app:littlegroupSH}.

\subsection{Helicity basis}
Here we discuss the helicity basis, which will help us write operators as the representation of the little group. We will also need the polarization vector ($\xi$) which is defined as $p.\xi=0$. We choose three polarization vectors orthogonal to $p_{\alpha\dot\alpha}$ in the spinor representation as 
\footnote{The covariant polarization given in eq\eqref{spin1IJ} maps to the component polarization(+,-,0) as, 
\begin{align*}
     \xi_{1 \alpha\dot\alpha}^+ \sim \zeta^1_\alpha \zeta^1_{\dot\alpha}, ~~ \xi_{1\alpha\dot\alpha}^- \sim \zeta^2_\alpha \zeta^2_{\dot\alpha}, ~~ \xi_{1\alpha\dot\alpha}^0 \sim \zeta^1_\alpha\zeta^1_{\dot\alpha}+\zeta^2_\alpha\zeta^1_{\dot\alpha},   
\end{align*}   }
\begin{align}\label{polarizationvectors}
  \xi_1^{\alpha\dot\alpha(+)}
  = \frac{\lambda^\alpha\tilde\lambda^{\dot\alpha}}{p}\,,
  \qquad
  \xi_1^{\alpha\dot\alpha(-)}
  = \frac{\rho^\alpha\tilde \rho^{\dot\alpha}}{p}\,,
  \qquad
  \xi_1^{\alpha\dot\alpha(0)}
  = \frac{\rho^\alpha\tilde\lambda^{\dot\alpha} + \lambda^\alpha\tilde \rho^{\dot\alpha}}{p}\,.
\end{align}
This representation of the polarization tensors are $GL(1,R)$ scaling invariant. The polarization vectors discussed above can be used to project the integer spin current in this helicity basis. The transformation of the polarization under the little group and helicity flipping operation is discussed in Appendix \ref{app:polarunderlittlegroup}.

For the case of half integer current operators,
the polarization spinors can be chosen to be as fundamental representation of the $SL(2,\mathbb R)_L$ and $SL(2,\mathbb R)_R$ as follows,
\begin{align}\label{polarizationspinors}
  \xi_{L\frac{1}{2}}^{\alpha(+)}
  = \frac{\lambda^\alpha}{\sqrt{p}},
  \qquad
  \xi_{L\frac{1}{2}}^{\alpha(-)}
  = \frac{\rho^\alpha}{\sqrt{p}},
  \qquad
  \xi_{R\frac{1}{2}}^{\dot\alpha(+)}
  = \frac{\tilde\lambda^{\dot\alpha}}{\sqrt{p}},
  \qquad
  \xi_{R\frac{1}{2}}^{\dot\alpha(-)}
  = \frac{\tilde\rho^{\dot\alpha}}{\sqrt{p}}.
\end{align}
This representation of polarization spinors have non-zero weight under $GL(1, \mathbb R)$ scaling.

Now, given a conserved symmetric traceless current $J^{\mu_1 \cdots \mu_s}$ $\sim$ $J_{(\alpha_1\alpha_2\cdots\alpha_s)(\dot\alpha_1\dot\alpha_2\cdots\dot\alpha_s)}$ of integer spin s, we can define 
  \begin{align}
       J_s^{+s}&= \xi_1^{\alpha_1\dot\alpha_1(+)}\xi_1^{\alpha_2\dot\alpha_2(+)}\cdots \xi_1^{\alpha_s\dot\alpha_s(+)} J_{(\alpha_1\alpha_2\cdots\alpha_s)(\dot\alpha_1\dot\alpha_2\cdots\dot\alpha_s)}~~,  \notag \\ &\vdots \notag \\
       J_s^{0}&= \xi_1^{\alpha_1\dot\alpha_1(0)}\xi_1^{\alpha_2\dot\alpha_2(0)}\cdots \xi_1^{\alpha_s\dot\alpha_s(0)} J_{(\alpha_1\alpha_2\cdots\alpha_s)(\dot\alpha_1\dot\alpha_2\cdots\dot\alpha_s)}~~, \notag \\  &\vdots \notag \\
      J_s^{-s}&= \xi_1^{\alpha_1\dot\alpha_1(-)}\xi_1^{\alpha_2\dot\alpha_2(-)}\cdots \xi_1^{\alpha_s\dot\alpha_s(-)} J_{(\alpha_1\alpha_2\cdots\alpha_s)(\dot\alpha_1\dot\alpha_2\cdots\dot\alpha_s)}~~,
  \end{align}
and similarly, given a current $J_{(\alpha_1\alpha_2\cdots\alpha_m)(\dot\alpha_1\dot\alpha_2\cdots\dot\alpha_n)}$ of half-integer spin s, i.e. $ \ | m-n\ | \in \mathbb{Z}^+$, we can define different helicity components as
  \begin{align}
       J_{s}^{+s} &= \xi_{\frac{1}{2}}^{\alpha_1(+)}\cdots\xi_{\frac{1}{2}}^{\alpha_m(+)} \xi_{\frac{1}{2}}^{\dot\alpha_1(+)}\cdots\xi_{\frac{1}{2}}^{\dot\alpha_n(+)} J_{(\alpha_1\cdots\alpha_m)(\dot\alpha_1\cdots\dot\alpha_n)}~~,  \notag \\
      J_{s}^{+(s-1)}&= \xi_{\frac{1}{2}}^{\alpha_1(-)}\xi_{\frac{1}{2}}^{\alpha_2(+)} \cdots\xi_{\frac{1}{2}}^{\alpha_m(+)} \xi_{\frac{1}{2}}^{\dot\alpha_1(-)}\xi_{\frac{1}{2}}^{\dot\alpha_2(+)}\cdots\xi_{\frac{1}{2}}^{\dot\alpha_n(+)} J_{(\alpha_1\cdots\alpha_m)(\dot\alpha_1\cdots\dot\alpha_n)}~~,  \notag \\
      &~~\vdots \notag \\
      J_{s}^{-s}&= \xi_{\frac{1}{2}}^{\alpha_1(-)} \cdots\xi_{\frac{1}{2}}^{\alpha_m(-)} \xi_{\frac{1}{2}}^{\dot\alpha_1(-)}\cdots\xi_{\frac{1}{2}}^{\dot\alpha_n(-)} J_{(\alpha_1\cdots\alpha_m)(\dot\alpha_1\cdots\dot\alpha_n)},
      \end{align}
where $J_s^{h}$ represents a spin-s current operator and h represents the different helicity components.
 
Note that we find it useful to normalize the above defined currents as to make the action of conformal generators to be simple,
 \begin{align}
    \hat J_s^{h} \equiv \frac{J^h_s}{p^{s}}~.
 \end{align}

\subsection{The Grassmannian framework}
  The way to introduce helicity basis Grassmannian is same as approached in sub-section \ref{subsec:covariantsympbigrassman}, just written in a component way. The details has been worked out in Appendix \ref{app:HBGrassframe}. 

  With the Grassmannian has been introduced, The $n$-point conformal correlator with helicities $h_1,h_2,\ldots,h_n$ is
represented as
\begin{align}\label{eq:grassinthelicitybasis}
  \boxed{
  \psi_n^{h_1,h_2,\ldots,h_n}
  \;=\;
  \int\frac{\mathcal{D}C\;\mathcal{D}\tilde C}
        {\Vol\bigl(GL(k)\bigr)\,\Vol\bigl(GL(2n-k)\bigr)}
  \;\delta\bigl( C\cdot\tilde\Lambda\bigr)\,
  \delta\bigl(\tilde C\cdot\Lambda\bigr)\,
  \delta\bigl(\tilde C\cdot\Omega\cdot C^T\bigr)\;
  \mathcal{F}^{h_1,\ldots,h_n}(C,\tilde C)
  },
\end{align}
where ${C}$ and $\tilde C$ as k-plane and (2n-k)-plane are denoted by,  
\begin{align}\label{CandCtilde2}
C_{iA} \equiv C =
\begin{pmatrix}
c_{11} & c_{12} & \cdots & c_{12n} \\
c_{21} & c_{22} & \cdots & c_{22n} \\
\vdots & \vdots & \vdots & \vdots \\
c_{k1} & c_{k2} & \cdots & c_{k2n}
\end{pmatrix}, ~~\tilde C_{iA} \equiv \tilde C =
\begin{pmatrix}
\tilde c_{11} & \tilde c_{12} & \cdots & \tilde c_{12n} \\
\tilde c_{21} & \tilde c_{22} & \cdots & \tilde c_{22n} \\
\vdots & \vdots & \vdots & \vdots \\
\tilde c_{(2n-k)1} & \tilde c_{(2n-k)2} & \cdots & \tilde c_{(2n-k)2n}
\end{pmatrix} 
\end{align}
and the metric on the sympletic manifold is denoted by 
 \begin{align}
 \Omega_{AB}=\Omega^{AB}= \begin{pmatrix}
        0&\mathbf 1_{n\times n} \\
        - \mathbf{1}_{n\times n}&0
    \end{pmatrix}.
\end{align}
  Before going to constraint $\mathcal{F(C,\tilde C)}$, we need to ask about the action of the little group symmetries on the Symplectic Grassmannian (C, $\tilde{C}$).
\subsection{Little group symmetry action on the Matrices $C$ and $\tilde{C}$}
\label{sec:notation}

We adopt the following column notation, which makes the symmetry
actions transparent:
\begin{align}
  C
  &=\bigl\{\,c_{1},\,c_{2},\,\ldots,\,c_{n},\;
     c_{n+1},\,\ldots,\,c_{2n}\,\bigr\}= \bigl\{\,\cund{1},\,\cund{2},\,\ldots,\,\cund{n},\;
             \cbar{1},\,\cbar{2},\,\ldots,\,\cbar{n}\,\bigr\},
  \\
  \tilde{C}
  &=\bigl\{\,\tilde c_{1},\,\tilde c_{2},\,\ldots,\,\tilde c_{n},\;
     \tilde c_{n+1},\,\ldots,\,\tilde c_{2n}\,\bigr\}= \bigl\{\,\cuts{1},\,\cuts{2},\,\ldots,\,\cuts{n},\;
             \ctil{1},\,\ctil{2},\,\ldots,\,\ctil{n}\,\bigr\},
\end{align}
where $\cund{i}\equiv c_i$, $\cbar{i}\equiv c_{n+i}$,
$\cuts{i}\equiv\tilde c_i$, $\ctil{i}\equiv\tilde c_{n+i}$.
The labels $1,2,\ldots,n$ index the external operators.

Under the action of the $GL(1,\mathbb R)$ and $SL(2,\mathbb R)$ on the external spinor helicity variables as given in Appendix \ref{app:littlegroupSH}, using eq\eqref{eq:finiteGL1gen} and eq\eqref{eq:finiteSL2gen}, we demand the following action on Grassmannian, 
 \begin{align}\label{eq:GL1action}
 GL(1,\mathbb{R})(G):\quad C \xrightarrow{}
  \begin{pmatrix}
      {\bar r_1}\cund{1}, & {\bar r_2}~\cund{2}, & \cdots& {\bar r_1}\cbar{1}, & {\bar r_2}\cbar{2},&\cdots 
  \end{pmatrix}, \notag \\
  \tilde C \xrightarrow{}
  \begin{pmatrix}
      \frac{1}{\bar r_1}\cuts{1} & \frac{1}{\bar r_2}\cuts{2} & \cdots& \frac{1}{\bar r_1}\ctil{1} & \frac{1}{\bar r_2}\ctil{2}&\cdots  
  \end{pmatrix}.
\end{align}
\begin{align}
  SL(2,\mathbb{R}):\quad (S^1_1=-S^2_2): C \xrightarrow{}
  \begin{pmatrix}
      {r_1}\cund{1} & {r_2}\cund{2} & \cdots& \frac{1}{ r_1}\cbar{1} & \frac{1}{r_2}\cbar{2}&\cdots 
  \end{pmatrix}, \notag \\
  \tilde C \xrightarrow{}
  \begin{pmatrix}
      {r_1}\cuts{1} & { r_2}\cuts{2} & \cdots& \frac{1}{ r_1}\ctil{1} & \frac{1}{ r_2}\ctil{2}&\cdots  
  \end{pmatrix},\label{eq:SL11action}
\end{align}
~~~~\text{For each external operator $i$:} 
\begin{align}
S^1_2\text{ (raising)}: &\quad \cund{i}\to\cbar{i},\quad \cuts{i}\to\ctil{i}, \notag\\
  S^2_1\text{ (lowering)}: &\quad \cbar{i}\to\cund{i},\quad \ctil{i}\to\cuts{i}.
  \label{eq:raisingloweringongrassmannian} 
\end{align}

Now we can proceed to find constraints on $\mathcal{F}^{h_1,h_2, \cdots h_n}(C,\tilde C)$ due to the above mentioned symmetry action on Grassmannian. 
\subsection{Constraints on $\mathcal{F}(C,\tilde{C})$}
\label{sec:constraints}

The function $\mathcal{F}^{h_1h_2\cdots h_n}(C,\tilde C)$ must satisfy the following
constraints, which arise from requiring the full integral
\eqref{eq:grassinthelicitybasis} to transform correctly under all symmetries.
\begin{enumerate}
\item \textbf{$GL(1,\mathbb{R})$ covariance:}
 Under the action of eq\eqref{eq:GL1action},  \begin{align}\label{eq:FconstraintGL1}
     \mathcal{F}^{h_1h_2\cdots h_n}(C,\tilde C)\;\longrightarrow\;(\bar r_1^{2h_1}\bar r_2^{2h_2}\cdots \bar r_n^{2 h_n})(\bar r_1^{4(n-k)}\bar r_2^{4(n-k)}\cdots \bar r_n^{4(n-k)})   \mathcal{F}^{h_1h_2\cdots h_n}(C,\tilde C). 
  \end{align}
  Please note that, since the polarization spinors of half-integer current has non trivial rescaling under GL(1,R) as discussed in eq\eqref{polarizationspinors}. So, if the $i^{th}$ operator is a half-integer spin current, then set $h_i= h_i$ or else if it is integer spin current, set $h_i=0$ in the above transformation.
\item \textbf{$S^1_1 = -S^2_2$ covariance (spin weights $h_i$)}: Under the action of eq\eqref{eq:SL11action} \footnote{One might suspect that we can write in the numerator lot of ansatz having the correct $SL^1_1$ demand, but the idea we will follow is that, we will start from all the correlator from the all + helicity and get all the other helicity correlators by helicity flipping operation\eqref{eq:helicityflipvectorgrass}, so that the full correlator is closed under the $SL(2,\mathbb R)$ action. },
\begin{align}\label{eq:FconstraintSL1}
    \mathcal{F}^{h_1h_2\cdots h_n}(C,\tilde C)
  \;\longrightarrow\;
  \frac{1}{r_1^{2h_1}r_2^{2h_2}\cdots r_n^{2h_n}}\,\mathcal{F}^{h_1h_2\cdots h_n}(C,\tilde C).
\end{align}



\item Full integral should be invariant under $GL$ transformation of the respective k-plane and (2n-k)-plane. Hence, under $GL(k)$ and $GL(2n-k)$ transformation i.e. 
$C\to g\cdot C$ and $\tilde C\to\tilde g\cdot\tilde C$:
\begin{align}\label{eq:FconstriantGLk}
   \mathcal{F}(C,\tilde C)
  \;\longrightarrow\;
  \det(g)^{2-k}\,\det(\tilde g)^{-2n+2+k}\,\mathcal{F}(C,\tilde C).    
\end{align}

\item $\mathcal{F}(C,\tilde C)$ is built purely from
$k$ and $2n-k$ dimensional minors of $C$ and $\tilde C$ respectively.
\item One can go from one helicity component to another in Grassmannian by using the helicity flipping operator eq(\eqref{eq:helicityflipvectorSH}) on $\mathcal{F}(C, \tilde C)$. Considering $\hat h$ on $i^{th}$ operator of integer spin,
         \begin{align}\label{eq:helicityflipvectorgrass}
             \mathcal F^{0} &= \mathcal F^{+}|_{\cbar{i} \xrightarrow{} \cund{i} } + \mathcal F^{+}|_{\ctil{i} \xrightarrow{} \cuts{i}} \,, \notag \\
             \mathcal F^{0} &= \mathcal F^{-}|_{\cund{i} \xrightarrow{} \cbar{i}} + \mathcal F^{-}|_{\cuts{i} \xrightarrow{} \ctil{i}} \,, \notag \\
             \mathcal F^{-} &= \mathcal F^{+}|_{\cbar{i} \xrightarrow{} \cund{i},~ \ctil{i} \xrightarrow{} \cuts{i}} \,, \notag \\
             \mathcal F^{+} &= \mathcal F^{-}|_{\cund{i} \xrightarrow{} \cbar{i},~ \cuts{i} \xrightarrow{} \ctil{i}} \, , \notag \\
             \mathcal F^{+} &= \mathcal F^{0}|_{\cund{i} \xrightarrow{} \cbar{i},~ \cuts{i} \xrightarrow{} \ctil{i}}\, , \notag \\
             \mathcal F^{-} &= \mathcal F^{0}|_{\cbar{i} \xrightarrow{} \cund{i},~ \ctil{i} \xrightarrow{} \cuts{i}}\,  .
         \end{align}
Similarly, can be obtained for half-integer spin current also.

\end{enumerate}

\subsection{Procedure for Constructing $\mathcal{F}(C,\tilde{C})$}
Please note that we will be focusing on k =n in this paper. One can still use k=2 where the integral over the Grassmannian will be trivial as can be observed in Appendix(\ref{app:countingofintegral}). In k=2, we have to write the $\mathcal F(C,\tilde C)$ containing minors of dimension $2$ and $2n-2$, one has to explore in this direction to see its usefulness. The systematic construction proceeds in three steps,

\textbf{ (1) $GL(1,\mathbb{R})$ invariant blocks:}  To form GL(1,R) invariant blocks we form a bilinear block made out of $n\times n$ minors of C and $\tilde C$, which has zero projective weight under GL(1,R) transformation for each operator.

\textbf{(2) $SL(2, \mathbb{R})$ invariant combinations:}  In order to form SL(2,R) invariant blocks one can use the above formed GL(1,R) invariant blocks and then act it with
$S^1_2|_i$ and $S^2_1|_i$ on each operators simultaneously and find linear
combinations of GL(1,R) blocks which are annihilated by the raising-lowering operations.  These will give the $SL(2,\mathbb{R})$ invariant
blocks $\Block_n$. `n` represents the no. of operators in the correlator.

\textbf{(3) Fix $GL(n)$ weight:}  The $SL(2,\mathbb{R})$ invariant block
$\Block_n$ goes in the \emph{denominator} of $\mathcal{F}(C,\tilde C)$. The numerator carries the minors with appropriate helicity
weights.  The power of $\Block_n$ in the denominator is fixed by requiring $\mathcal{F}(C,\tilde C)\to\det(g)^{2-n}\det(\tilde g)^{2-n}$ $\mathcal{F}(C,\tilde C)$ under GL(n) transformation of C and $\tilde C$.

\subsection{Examples}
 We start with examples of bootstrapping the $\mathcal{F}(C, \tilde{C})$ given the above rules. We will work out the example for two points correlators in detail and will only layout the examples for three point correlators. 
\subsubsection{Two points (n= 2, k= 2) }
Following the steps as mentioned above, we have to form $\mathbb {GL}(1,\mathbb R)$ and $\mathbb {SL}(2,\mathbb R)$  invariant blocks. 

\textbf{Step 1:} The $GL(1,\mathbb{R})$ invariant blocks split into two sets.
\begin{align}
    & \text{\textbf{Set 1} (mixed):}  ~(\cbar{1}\cbar{2})(\cuts{1}\cuts{2}),\quad
  (\cund{1} \cund{2})(\ctil{1}\ctil{2}),\quad
  (\cbar{1}\cund{2})(\cuts{1}\ctil{2}),\quad
  (\cund{1}\cbar{2})(\ctil{1}\cuts{2}). \notag \\
& \text{\textbf{Set 2} (diagonal):}~ (\cbar{1}\cund{1})(\cuts{1}\ctil{1}),\qquad
  (\cbar{2}\cund{2})(\cuts{2}\ctil{2}).
\end{align}

\textbf{Step 2:} Demanding annihilation by $S^1_2|_i$ and $S^2_1|_i$ for $i=1,2$
individually giving three $SL(2,\mathbb{R})$ invariant block:
\begin{align}\label{eq:Block2}
  & \Block_2^{(12)}
  \;=\;
  (\cbar{1}\cbar{2})(\cuts{1}\cuts{2})
  -(\cbar{1}\cund{2})(\cuts{1}\ctil{2})
  -(\cund{1}\cbar{2})(\ctil{1}\cuts{2})
  +(\cund{1}\cund{2})(\ctil{1}\ctil{2}), \notag \\
  & \Block_2^{(11)}
  \;=\;
  (\cbar{1}\cund{1})(\cuts{1}\ctil{1}), 
  ~~\Block_2^{(22)}
  \;=\;
  (\cbar{2}\cund{2})(\cuts{2}\ctil{2}).
\end{align}

The constrained $C$ and $\tilde C$ i.e. $\delta(\tilde C\cdot\Omega\cdot C^T)$ and in a particular gauge is given by,
\begin{equation}
C=\begin{pmatrix}1&0&c_{11}&c_{12}\\0&1&c_{21}&c_{22}\end{pmatrix},\qquad
  \tilde C=\begin{pmatrix}1&0&c_{11}&c_{21}\\0&1&c_{12}&c_{22}\end{pmatrix}.
\end{equation}
Using the plucker relation with the above expression for C and $\tilde C$ one can show that, $\Block_2^{(12)} = \Block_2^{(11)} = \Block_2^{22} = -2c_{12}c_{21}$. Hence, there is only one $SL(2,\mathbb R)$ invariant block. 
\begin{align}
    \Block_2^{(12)}= \Block_2^{(11)}= \Block_2^{(22)}.
\end{align}

\textbf{Step 3:} Now, demanding the correct helicity weight and GL(k=2) weight of the correlator using eq\eqref{eq:FconstraintGL1}, eq\eqref{eq:FconstraintSL1} and eq\eqref{eq:FconstriantGLk}, we fix the numerator and the denominator of various $\mathcal{F}(C, \tilde{C})$.   
          \begin{align}\label{eq:fcjsjs}
              n=2, h_1=h_2= +s, s \in \mathbb{Z}^+ :~~
              F^{+s,+s}_{J_sJ_s}(C,\tilde C) = \frac{(\cbar1 \cbar2)^{s}(\ctil1 \ctil2)^{s}}{(\Block_2^{(12)})^s}
         \end{align}
         \begin{align}\label{eq:fcjhalfjhalf}
              n=2, h_1=h_2= +\frac{1}{2}, s \in \mathbb{Z}^+ :~~
              F^{+\frac{1}{2},+\frac{1}{2}}_{J_\frac{1}{2}J_\frac{1}{2}}(C,\tilde C) = \frac{(\cbar1 \cbar2)(\ctil2 \cuts2)}{\Block_2^{(12)}}
         \end{align}   
We can use the helicity flipping operator eq\eqref{eq:helicityflipvectorgrass} in Grassmannian to get the two point function for other helicities.  
For example,
    \begin{align}
               \mathcal F^{+(s-1),+(s-1)}_{\hat J_s\hat J_s}(C,\tilde C) &= \frac{[(\cbar1 \cbar2)(\ctil1 \ctil2)]^{s-1}}{(\Block_2^{(12)})^{s-1}}\frac{[(\cbar{1}\cbar{2})(\cuts{1}\cuts{2})
  +(\cund{1}\cund{2})(\ctil{1}\ctil{2})
  +(\cbar{1}\cund{2})(\cuts{1}\ctil{2})
  +(\cund{1}\cbar{2})(\ctil{1}\cuts{2})]}{\Block_2^{(12)}}\notag\\
  &= \frac{[(\cbar1 \cbar2)(\ctil1 \ctil2)]^{s-1}}{(\Block_2^{(12)})^{s-1}} \mathcal{F}_{\hat J_1\hat J_1}^{0,0}(C,\tilde C)
         \end{align}

One can check the validity of the above correlators $\mathcal F(C,\tilde C)$, by matching it with different helicity components of the covariant answers given in sec \ref{sec:2pointcovariantgrass} by using the map for $J^{(IJ)}$ to get $J^{(11)}\sim J^+, J^{(22)}\sim J^-$ and $J^{(12)}\sim J^0$. One can also convert the above expressions in helicity basis to helicity basis in momentum space to match with the results given in eq (\ref{sec:2pointSH}).

\subsubsection{Three points}
    A similar procedure can be followed, as in the above section to get the SL(2,R) invariant blocks, 
      \begin{align}
           \Block_3^{(12)} &= (\cbar{1}\cund{1}\cbar{2})(\cuts{1}\ctil{1}\cuts{2})
            -(\cbar{1}\cund{1}\cund{2})(\cuts{1}\ctil{1}\ctil{2}),\notag\quad \Block_3^{(13)}= (\cbar{1}\cund{1}\cbar{3})(\cuts{1}\ctil{1}\cuts{3})
            -(\cbar{1}\cund{1}\cund{3})(\cuts{1}\ctil{1}\ctil{3}),\\
  \Block_3^{(23)} &= (\cbar{2}\cund{2}\cbar{3})(\cuts{2}\ctil{2}\cuts{3})
            -(\cbar{2}\cund{2}\cund{3})(\cuts{2}\ctil{2}\ctil{3}).
      \end{align}
All the three SL(2,R) invariant blocks are same using plucker relations.

 Now, we present some of the examples of the three point correlators in the (+) helicity: 
 \begin{tcolorbox}[
  enhanced,
  colback=white,
  colframe=blue!50!black,
  arc=3pt,
  boxrule=1pt,
  left=0pt, right=0pt, top=0pt, bottom=0pt
]
\renewcommand{\arraystretch}{3.2}
\begin{tabular}{
  !{\color{blue!50!black}\vrule width 1pt}
  F
  !{\color{blue!50!black}\vrule}
  R
  !{\color{blue!50!black}\vrule width 1pt}
}
\rowcolor{blue!10}
\textbf{3-point} & \textbf{Results}
\\
\hline
\rowcolor{gray!8}
\mathcal{F}_{\hat{O}_2\hat{O}_2\hat{O}_2}^{0,0,0}(C,\tilde{C})
&
\dfrac{1}{\Block_3^{(12)}} Sgn(\Block_3^{(12)}) 
\vspace{1pt}
\\
\hline
\mathcal{F}_{\hat{J}\hat{O}_2\hat{O}_2}^{+1,0,0}(C,\tilde{C})
&
\frac{(\cbar1(\Block_2^{(23)}) \ctil1)}{(\Block_3^{(12)})^{2}} Sgn(\Block_3^{(12)}) 
\vspace{1pt}
\\
\hline
\rowcolor{gray!8}
\mathcal{F}_{\hat{T}\hat{O}_2\hat{O}_2}^{+2,0,0}(C,\tilde{C})
&
\frac{(\cbar1(\Block_2) \ctil1)^2}{(\Block_3^{(12)})^{3}} Sgn(\Block_3^{(12)}) 
\vspace{1pt}
\\
\hline
\mathcal{F}_{\hat{J}\hat{J}\hat{O}_2}^{+1,+1,0}(C,\tilde{C})
&
\frac{(\cbar1\cbar2\cbar3)(\ctil1\ctil2\cuts3)-(\cbar1\cbar2\cund3)(\ctil1\ctil2\ctil3)}{(\Block_3^{(12)})^{2}} Sgn(\Block_3^{(12)}) 
\vspace{1pt}
\\
\hline
\rowcolor{gray!8}
\mathcal{F}_{\hat{T}\hat{T}\hat{O}_2}^{+2,+2,0}(C,\tilde{C})
&
\frac{((\cbar1\cbar2\cbar3)(\ctil1\ctil2\cuts3)-(\cbar1\cbar2\cund3)(\ctil1\ctil2\ctil3))^2}{(\Block_3^{(12)})^{3}} Sgn(\Block_3^{(12)}) 
 \vspace{1pt}     
\\
\hline 
\mathcal{F}_{\hat{J}\hat{J}\hat{J}}^{+1,+1,+1}(C,\tilde{C})
&
\begin{array}[t]{l}
  \bigg(\alpha_1\,\dfrac{(\cbar{1}\cbar{2}\cbar{3})(\ctil{1}\ctil{2}\ctil{3})}
                  {(\Block_3^{(12)})^{2}}
  +\;\beta_1\,
  \dfrac{(\cbar{1}\,\Block_2^{(23)}\,\ctil{1})\,
         (\cbar{2}\,\Block_2^{(31)}\,\ctil{2})\,
         (\cbar{3}\,\Block_2^{(12)}\,\ctil{3})}
        {(\Block_3^{(12)})^{4}} \bigg)  Sgn(\Block_3^{(12)}) 
\\[8pt] 
\end{array}
\\
\hline
\end{tabular}
\end{tcolorbox}

\begin{tcolorbox}[
  enhanced,
  colback=white,
  colframe=blue!50!black,
  arc=3pt,
  boxrule=1pt,
  left=0pt, right=0pt, top=0pt, bottom=0pt
]
\renewcommand{\arraystretch}{3.2}
\begin{tabular}{
  !{\color{blue!50!black}\vrule width 1pt}
  F
  !{\color{blue!50!black}\vrule}
  R
  !{\color{blue!50!black}\vrule width 1pt}
}
\rowcolor{gray!8}
\mathcal{F}_{\hat{T}\hat{T}\hat{T}}^{+2,+2,+2}(C,\tilde{C})
&
\begin{array}[t]{l}
  \alpha_2\,
  \dfrac{\bigl[(\cbar{1}\cbar{2}\cbar{3})(\ctil{1}\ctil{2}\ctil{3})\bigr]^{2}}
        {(\Block_3^{(12)})^{3}} Sgn(\Block_3^{(12)}) 
    \\[5pt]
  +\;\beta_2\,
  \dfrac{\bigl[(\cbar{1}\,\Block_2^{(23)}\,\ctil{1})\,
               (\cbar{2}\,\Block_2^{(31)}\,\ctil{2})\,
               (\cbar{3}\,\Block_2^{(12)}\,\ctil{3})\bigr]^{2}}
        {(\Block_3^{(12)})^{7}} Sgn(\Block_3^{(12)}) 
  \\[5pt]
  +\;\gamma_2\,
  \dfrac{(\cbar{1}\cbar{2}\cbar{3})(\ctil{1}\ctil{2}\ctil{3})}
        {(\Block_3^{(12)})^{2}}
  \dfrac{(\cbar{1}\,\Block_2^{(23)}\,\ctil{1})\,
         (\cbar{2}\,\Block_2^{(31)}\,\ctil{2})\,
         (\cbar{3}\,\Block_2^{(12)}\,\ctil{3})}
        {(\Block_3^{(12)})^{3}} Sgn(\Block_3^{(12)}) 
\\[8pt]
\end{array}
\\
\hline
\end{tabular}
\end{tcolorbox}

where,
\begin{align*}
    (i~\Block_2^{(kl)}~j) \equiv (i\cbar{k}\cbar{l})(\cuts{k}\cuts{l}j)
  -(i\cbar{k}\cund{l})(\cuts{k}\ctil{l}j)
  -(i\cund{k}\cbar{l})(\ctil{k}\cuts{l}j)
  +(i\cund{k}\cund{l})(\ctil{k}\ctil{l}j).
\end{align*}
One can check the validity of the above correlators $\mathcal F(C,\tilde C)$, by matching it with different helicity components of the covariant answers given in sec \ref{sec:3pointcovariantgrass} or also by converting these answers to momentum space in the helicity basis and match with the spinor helicity answers given in sec \ref{sec:3pointSH}.
We can use the helicity flipping operation in eq\eqref{eq:helicityflipvectorgrass} to get the other components.

\section{Route to Twistors from Grassmannian}\label{sec:Twistor}
    Given the simplicity of the Grassmannian constructions, we now seek a twistorial description. Twistors are usually the best set of variables to use in the study of conformal field theories \cite{Penrose:1973vca,Penrose:1986ca,Ward:1990vs}, as they make conformal invariance manifest. Witten introduced the half-Fourier transform from spinor helicity to twistor space in the context of massless scattering amplitudes \cite{Witten:2003nn}. The basic idea is to transform the on shell spinor variables pair $(\lambda,\tilde{\lambda})$ to $Z=(\lambda,\mu)$ or $W=(\tilde{\mu},\tilde{\lambda})$ where, $\mu$ and $\tilde{\mu}$ are Fourier conjugate to $\tilde{\lambda}$ and $\lambda$ respectively. 

    In our off-shell construction, we can perform a similar operation and get twistors in two in-equivalent ways. 
    These two in-equivalent ways are, 
        \begin{align}
            \{\lambda_\alpha, \rho_\alpha, \tilde\lambda_{\dot\alpha}, \tilde\rho_{\dot\alpha} \} &\xrightarrow{h.F.T}  W_A= \{ \mu^{\alpha}\oplus \tilde\lambda_{\dot\alpha}   \} ~~and~~ Z^{A}= \{ \lambda_{\alpha}\oplus \tilde\mu^{\dot\alpha}   \}    
            \notag \\ 
            \{\lambda_\alpha^I,\tilde\lambda_{\dot\alpha}^I\} &\xrightarrow{h.F.T}  W^{(I)}_A= \{ \mu^{\alpha(I)}\oplus \tilde\lambda_{\dot\alpha}^{(I)}   \} ~~or~~ Z^{(I)A}= \{ \lambda_{\alpha}^{(I)}\oplus \tilde\mu^{\dot\alpha(I)}   \},
        \end{align}
    where, the index (I) and (A) represents the fundamental of $SL(2,\mathbb R)$ and the double cover of the Conf(2,2) which is $SL(4,\mathbb R)$ respectively. In the first approach $\rho $ and $\tilde \rho$ is half-Fourier transformed to give a pair of Twistors called Ambi-Twistors. In this approach the $SL(2,\mathbb R)$ covariance is not manifest. In the second approach either $\lambda^I$ or $\tilde\lambda^I$ is half-Fourier transformed to get $SL(2,\mathbb{R})$ Twistors. In the latter approach the SL(2,R) covariance is preserved as the twistors carry the SL(2,R) index with them. 

    The above discussion suggests that the former Twistor is best suited to the helicity basis Grassmannian, and the latter is best suited to SL(2,R) covariant Grassmannian. So with this, we directly present the examples of conformal correlators in the language of both twistors by performing the half Fourier transform here. Further the twistor construction and more results will be discussed in detail in future in the upcoming paper \cite{XYZ:2026new4}. 
    
    \subsection{Ambi-Twistors}
        We present here the results of correlators following the covariant Grassmannian results in sec \ref{sec:HBgrassframework},
        \begin{align}
            \langle0|O_2(Z_1, W_1) O_2(Z_2, W_2)|0\rangle = \delta(Z_1\cdot W_1)\delta(Z_2\cdot W_2)\delta(Z_1\cdot W_2)\delta(Z_2\cdot W_1), 
        \end{align}
         \begin{align}
            \langle0|J^+(Z_1, W_1) J^+(Z_2, W_2)|0\rangle = \delta(Z_1\cdot W_1)\delta(Z_2\cdot W_2) Sgn(Z_1\cdot W_2) Sgn(Z_2\cdot W_1), 
        \end{align}
        \begin{align}
            \langle0|&O_2(Z_1, W_1) O_2(Z_2, W_2) O_2(Z_3, W_3)|0\rangle \notag \\ 
            &=  \frac{\delta(Z_1\!\cdot\! W_1)\,
\delta(Z_2\!\cdot\! W_2)\,
\delta(Z_3\!\cdot\! W_3)}
{|(Z_1\!\cdot\! W_2)\,
(Z_2\!\cdot\! W_3)\,
(Z_3\!\cdot\! W_1)|} ~ \log(Z_1\cdot W_2\, Z_2\cdot W_3\, Z_3\cdot W_1) \delta(1+\tau) ,
        \end{align}
where $\tau= \frac{Z_1\cdot W_3 Z_3\cdot W_2 Z_2\cdot W_1}{Z_1\cdot W_2 Z_2\cdot W_3 Z_3\cdot W_1}$.
    We can observe one common thing in all the examples above, the Ambi-Twistor lives in a quadric space of dual Twistors obtained by half Fourier transform, i.e. $Z\cdot W= Z_i^{(A)} W_{i(A)}=0$. This is in alignment with the definition of ambitwistor space given in \cite{Adamo:2016rtr}. In that paper, they were only able to compute the scalar two point and spin-half 2 point in the AdS$_5$ bulk using ambitwistor to get the boundary correlator in the position space. Here we have directly the results in ambitwistor beyond two point and also for spinning correlators. Some of the work in Ambitwistor has been done in the literature \cite{Mason:2005zm,Mason:2013sva,Geyer:2014fka}.  Further results will be discussed in detail in future. 

     \subsection{$SL(2,\mathbb R)$ Twistors}
        We present here the results of correlators following the $SL(2,\mathbb R)$ Grassmannian results in sec \ref{sec:Covgrassframe},
        \begin{align}
            \langle0|O_2(Z_1^{(I)}) O_2(Z_2^{(I)})|0\rangle = \int \frac{d\tilde C}{Vol(GL(2))} \delta^{2,4}[ \tilde C\cdot Z^{A}]~ , 
        \end{align}    
        \begin{align}
            \langle0|O_2(Z_1^{(I)}) O_2(Z_2^{(I)})O_2(Z_3^{(I)})|0\rangle 
            = \int \frac{d\tilde C}{Vol(GL(3))} \delta^{3,4}[ \tilde C\cdot Z^{A}]~ \frac{1}{\Block(C,\tilde C)}\bigg|_{C=\tilde C_\perp}.
        \end{align}
 where, $\tilde C\cdot Z^A = \tilde C_{ij,(I)} (\Omega^{jk})^{IJ} Z_{k(J)}^{A}$ and $\tilde C_{\perp} = C$ is obtained by solving $\tilde C\cdot\Omega\cdot C = 0$ for C by choosing a particular GL(n) gauge in terms of $\tilde C$. It is interesting to note that the $SL(2,R)$ Twistor $Z$ are planes orthogonal to  $\tilde C$ planes, as can be observed from the delta function in the integrand. Further discussion on Twistor along with spinning correlators will appear in future publication.      

\section{Discussion}\label{sec:discussion}
In this paper we introduced a symplectic bi-Grassmannian formulation of four-dimensional
CFT Wightman correlators. Working in Klein space and using off-shell spinor-helicity
variables, we showed that correlators of \(\Delta=2\) scalars, spin-\(\frac12\) operators,
conserved currents and stress tensors can be organised in a little-group covariant way.
The central object is a pair of Grassmannian matrices \(C\) and \(\widetilde C\), constrained
by symplectic orthogonality and by their alignment with the external kinematic data. These
constraints make momentum conservation and conformal invariance geometric.

We tested the construction on two- and three-point functions. The resulting
Grassmannian expressions reproduce the corresponding off-shell spinor-helicity answers
and organise the tensor structures in terms of simple minors of \(C\) and \(\widetilde C\).
As a non-trivial check, the framework reproduces the full set of independent conformally
invariant structures of \(\langle JJJ\rangle\) and \(\langle TTT\rangle\) in CFT\(_4\).
The formalism also makes transparent a simple double-copy relation between the
Yang--Mills current correlator and the Einstein-gravity stress-tensor correlator.

There are several natural directions for future work. First, it would be important to
understand the twistor-space origin of the symplectic bi-Grassmannian. Since the present
construction is formulated in Klein space using real spinor variables, it should admit a
natural interpretation in terms of twistor or ambitwistor geometry\cite{Adamo:2016rtr}. This may clarify the
role of the two Grassmannian matrices and their relation to Penrose-transform
representations of CFT correlators.

A second direction is the extension to AdS\(_5\) amplitudes\cite{Alday:2021odx, Rastelli:2016nze} and Witten diagrams. The
three-point structures studied here already know about Yang--Mills, \(F^3\), Einstein,
Gauss--Bonnet and Weyl-cubed interactions in AdS\(_5\). It would be interesting to
formulate higher-point AdS\(_5\) exchange amplitudes directly in the symplectic
bi-Grassmannian language and to understand their flat-space limits.

Third, the present construction should be extended to genuine higher-point CFT
correlators. At four points and beyond, correlators contain dynamical information such as
OPE data, exchange singularities and crossing constraints. Understanding how these data
are encoded in the Grassmannian integral may lead to new recursion or factorisation
formulae for CFT Wightman functions.

Finally, the geometry itself deserves further study. The symplectic orthogonality
constraints suggest a close connection with the Lagrangian Grassmannian, while the
appearance of simple minor structures raises the possibility of a positive-Grassmannian or
amplituhedron-like interpretation for CFT correlators. A supersymmetric extension is also
natural: by adding suitable Grassmann variables, one may hope to construct a
super-bi-Grassmannian that packages scalar, fermion, current and stress-tensor correlators
into unified supercorrelators.

\section*{Acknowledgment}
AB acknowledges a UGC-JRF fellowship. D K.S. would like to thank Amit Suthar for discussions. We would like to thank  A. Rao for useful discussion and taking part at the initial stages of the project. We would especially like to acknowledge our debt to the
people of India for their steady support of research in basic sciences.  
\appendix

\section{Helicity-basis Grassmannian}
\label{app:helicity-basis-grassmannian}
\subsection{Little Group Generators in SH}\label{app:littlegroupSH}
   The differential representation of $GL(1,\mathbb{R})$ generator acting on the arbitrary function of spinor variables is given by:
\begin{align}
  G
  \;=\;
  \lambda\cdot\frac{\partial}{\partial\lambda}
  -\tilde\rho\cdot\frac{\partial}{\partial\tilde \rho}
  +\rho\cdot\frac{\partial}{\partial\rho}
  -\tilde\lambda\cdot\frac{\partial}{\partial\tilde\lambda}\,.
  \label{eq:GL1gen}
\end{align}

The three $SL(2,\mathbb{R})$ generators acting on the arbitrary function of spinor helicity variables is:
\begin{align}\label{eq:SIJgen}
  &S^1_1 = -S^2_2
  \;=\;
  \frac{1}{2}\!\left(
    \lambda\cdot\frac{\partial}{\partial\lambda}
    -\tilde \rho \cdot\frac{\partial}{\partial\tilde\rho}
    -\rho\cdot\frac{\partial}{\partial\rho}
    +\tilde\lambda\cdot\frac{\partial}{\partial\tilde\lambda}
  \right),
  \notag\\[6pt]
  S^1_2
  &\;=\;
  \left(\lambda\cdot\frac{\partial}{\partial\rho}
        +\tilde\lambda\cdot\frac{\partial}{\partial\tilde \rho}\right), \qquad
  S^2_1 
  \;=\;
  \left(\rho\cdot\frac{\partial}{\partial\lambda}
        +\tilde \rho\cdot\frac{\partial}{\partial\tilde\lambda}\right),
\end{align}
The algebra followed by the above generators is:
\begin{align}
    & [G, S^I_J]=0 \notag \\
    & [S^1_1, S^1_2]= + S^1_2, \qquad [S^1_1, S^2_1]= - S^2_1, \qquad [S_2^1,S^2_1]= 2 S^1_1 
\end{align}
Observing the algebra $sl(2,\mathbb R)$, one can see that it is isomorphic to $su(2)$ algebra. From which we can identify that $S_1^1$ is the Cartan generator, $S^1_2$ and $S_1^2$ are the raising and lowering operator. One can check that the action of these differential generators is trivial on eq\eqref{eq:momentumSH}. 

The infinitesimal matrix transformation at the level of spinor helicity variables is obtained by,

\textbf{$GL(1,\mathbb R)$}
\begin{align}\label{eq:finiteGL1gen}
    G: \lambda \xrightarrow{} \bar r \lambda, ~~\tilde\rho\xrightarrow{} \frac{1}{\bar r} \tilde\rho, ~~\rho \xrightarrow{} \bar r \rho,~~ \tilde\lambda\xrightarrow{} \frac{1}{\bar r} \tilde\lambda. 
\end{align}
and
\textbf{$SL(2,\mathbb R)$}
\begin{align}\label{eq:finiteSL2gen}
    S^1_1=-&S^2_2: \lambda \xrightarrow{} \frac{1}{r} \lambda, ~~\tilde\rho\xrightarrow{} r \tilde\rho, ~~\rho \xrightarrow{}  r \rho,~~ \tilde\lambda\xrightarrow{} \frac{1}{r} \tilde\lambda \notag \\
  & S^1_2: \rho \xrightarrow{} \lambda, ~~\tilde\rho\xrightarrow{} \tilde\lambda\notag \\
  & S^2_1: \lambda \xrightarrow{} \rho, ~~\tilde\lambda\xrightarrow{} \tilde\rho
\end{align}
\subsection{Polarization basis under little group}\label{app:polarunderlittlegroup}
We now describe how polarization tensors transforms under the little group transformations.

Under the $S^1_1$ action \eqref{eq:finiteSL2gen}, the polarization vector eq\eqref{polarizationvectors} scales with weight,
\begin{align}
    \epsilon_1^{(+)} \xrightarrow{} \frac{1}{r^2} \epsilon_1^{(+)},~~ \epsilon_1^{(-)} \xrightarrow{} {r^2} \epsilon_1^{(-)},~~ \epsilon_1^{(0)} \xrightarrow{} r^0 \epsilon_1^{(0)}.
\end{align}
Under the $S^1_1$ and G action, the polarization spinors eq\eqref{polarizationspinors} scale with weight,
\begin{align}
    &S^1_1: ~\epsilon_{L\frac{1}{2}}^{(+)} \xrightarrow{} \frac{1}{r} \epsilon_{L\frac{1}{2}}^{(+)},~~ \epsilon_{L\frac{1}{2}}^{(-)} \xrightarrow{} {r} ~\epsilon_{L\frac{1}{2}}^{(-)},~~ \epsilon_{R\frac{1}{2}}^{(+)} \xrightarrow{} \frac{1}{r} ~\epsilon_{R\frac{1}{2}}^{(+)},~~\epsilon_{R\frac{1}{2}}^{(-)} \xrightarrow{} {r} ~\epsilon_{R\frac{1}{2}}^{(-)} \notag \\ 
    & G: \epsilon_{L\frac{1}{2}}^{(+)} \xrightarrow{} {\bar r} \epsilon_{L\frac{1}{2}}^{(+)},~~ \epsilon_{L\frac{1}{2}}^{(-)} \xrightarrow{} {\bar r} ~\epsilon_{L\frac{1}{2}}^{(-)},~~\epsilon_{R\frac{1}{2}}^{(-)} \xrightarrow{} \frac{1}{\bar r} ~\epsilon_{R\frac{1}{2}}^{(-)},~~ \epsilon_{R\frac{1}{2}}^{(+)} \xrightarrow{} \frac{1}{\bar r} ~\epsilon_{R\frac{1}{2}}^{(+)}  .
\end{align}

  One can observe in eq \eqref{polarizationvectors} and eq \eqref{polarizationspinors}  that one can go from one basis to other by defining the helicity flipping operator\footnote{Please note that, while applying the helicity flipping operator we don't apply it on momentum p or any other SL(2,R) invariant object.} ($\hat h$) using the raising $S^1_2$ and lowering operator $S_1^2$ given in eq \eqref{eq:finiteSL2gen} as,
         \begin{align}\label{eq:helicityflipvectorSH}
             \xi_{1}^{\alpha\dot\alpha(0)} &= \xi_{1}^{\alpha\dot\alpha(+)}|_{\lambda \xrightarrow{} \rho} + \xi_{1}^{\alpha\dot\alpha(+)}|_{\tilde\lambda \xrightarrow{} \tilde\rho} \,, \notag \\
             \xi_{1}^{\alpha\dot\alpha(0)} &= \xi_{1}^{\alpha\dot\alpha(-)}|_{\rho \xrightarrow{} \lambda} + \xi_{1}^{\alpha\dot\alpha(-)}|_{\tilde\rho \xrightarrow{} \tilde\lambda} \,, \notag \\
             \xi_{1}^{\alpha\dot\alpha(-)} &= \xi_{1}^{\alpha\dot\alpha(+)}|_{\lambda \xrightarrow{} \rho, \tilde\lambda \xrightarrow{} \tilde\rho} \,, \notag \\
             \xi_{1}^{\alpha\dot\alpha(+)} &= \xi_{1}^{\alpha\dot\alpha(-)}|_{\rho \xrightarrow{} \lambda, \tilde\rho \xrightarrow{} \tilde\lambda} .
         \end{align}
\begin{align}\label{eq:helicityflipspinorSH}
             \xi_{L\frac{1}{2}}^{\alpha(+)} \xleftrightarrow{\lambda \xleftrightarrow{} \rho} \xi_{L\frac{1}{2}}^{\alpha(-)} , \notag \\
             \xi_{R\frac{1}{2}}^{\dot\alpha(+)} \xleftrightarrow{\tilde\lambda \xleftrightarrow{} \tilde\rho} \xi_{R\frac{1}{2}}^{\dot\alpha(-)} .
         \end{align}

\subsection{The Grassmannian Framework}\label{app:HBGrassframe}
We start with the momentum conservation equation,
\begin{align}\label{momentumconservation2}
    \sum_{i=1}^n p_{i{\alpha\dot\alpha}} &= \lambda_{1\alpha}\tilde\rho_{1\dot\alpha}-\rho_{1\alpha}\tilde\lambda_{1\dot\alpha}+\lambda_{2\alpha}\tilde\rho_{2\dot\alpha}-\rho_{2\alpha}\tilde\lambda_{2\dot\alpha}+\cdots \lambda_{n\alpha}\tilde\rho_{n\dot\alpha}-\rho_{n\alpha}\tilde\lambda_{n\dot\alpha}\notag \\
    &= \Lambda_\alpha^A\Omega_{AB}\tilde\Lambda^B_{\dot\alpha}=0,
\end{align}
where
\begin{align}
     \Lambda_\alpha^A = \begin{pmatrix}
\lambda_{1\alpha}\\ \lambda_{2\alpha}\\ \vdots\\ \lambda_n\\ \rho_{1\alpha}\\ \rho_{2\alpha}\\ \vdots\\ \rho_{n\alpha} 
    \end{pmatrix}~~~ ,~~~ \tilde\Lambda_{\dot\alpha}^A = \begin{pmatrix}
\tilde\lambda_{1\dot\alpha}\\ \tilde\lambda_{2\dot\alpha}\\ \vdots\\ \tilde\lambda_{n\dot\alpha}\\ \tilde\rho_{1\dot\alpha}\\ \tilde\rho_{2\dot\alpha}\\ \vdots\\ \tilde\rho_{n\dot\alpha} 
    \end{pmatrix} ~~and~~ \Omega_{AB}=\Omega^{AB}= \begin{pmatrix}
        0&\mathbf 1_{n\times n} \\
        - \mathbf{1}_{n\times n}&0
    \end{pmatrix}.
\end{align}
Hence, the external data of momenta live on a 2n dimensional symplectic manifold.

Now we define the $k\times 2n$ and $(2n-k)\times 2n$ dimensional matrices ${C}$ and $\tilde C$ as k-plane and (2n-k)-plane that respectively contain 2-planes, $\Lambda$ and $\tilde{\Lambda}$:  
\begin{align}\label{CandCtilde2}
C_{iA} \equiv C =
\begin{pmatrix}
c_{11} & c_{12} & \cdots & c_{1~2n} \\
c_{21} & c_{22} & \cdots & c_{2~2n} \\
\vdots & \vdots & \vdots & \vdots \\
c_{k1} & c_{k2} & \cdots & c_{k~2n}
\end{pmatrix}, ~~\tilde C_{iA} \equiv \tilde C =
\begin{pmatrix}
\tilde c_{11} & \tilde c_{12} & \cdots & \tilde c_{1~2n} \\
\tilde c_{21} & \tilde c_{22} & \cdots & \tilde c_{2~2n} \\
\vdots & \vdots & \vdots & \vdots \\
\tilde c_{(2n-k)1} & \tilde c_{(2n-k)2} & \cdots & \tilde c_{(2n-k)~2n}
\end{pmatrix}.
\end{align}

We further impose the constraint that C and $\tilde C$ are orthogonal,
\begin{align}\label{CdotCtilde2}
    \tilde{C}_{iA}\Omega^{AB}C^T_{Bj}= \tilde{C}\cdot\Omega\cdot C^T=0,
\end{align}
along with the conditions,
\begin{align}\label{CdotLambda2}
   &\tilde{C}_{iA}\Lambda_{\alpha}^{A}=0,\notag\\
   &C_{iA}\tilde{\Lambda}^{A}_{\dot\alpha}=0.
\end{align}
Geometrically one can understand the validity of the above equation, from the above discussions we can observe that C contain $\Lambda$ and $\tilde C$ contains $\tilde \Lambda$ and $C \perp \tilde C $, hence $\tilde C \perp \Lambda$ and $ C \perp \tilde\Lambda$.
It is easy to show that \eqref{CdotCtilde2} and \eqref{CdotLambda2} imply momentum conservation \eqref{momentumconservation2}. With the above at hand, we are ready to define the correlators on Symplectic Grassmannian.

\subsection{Counting of Integrals}\label{app:countingofintegral}

\paragraph{Integrals over $\tilde C$.}
The $\tilde C$ matrix has $(2n-k)\times 2n$ entries.  The constraints
remove:
\begin{equation}
  (2n-k)\cdot 2n - (2n-k)^2 - (2n-k)k
  = (2n-k)\bigl(2n - 2n + k - k\bigr) = 0.
\end{equation}
So the $\tilde C$ integrals can be completely done.

\paragraph{Integrals over $C$.}
The $C$ matrix has $k\times 2n$ entries.  The constraints remove:
\begin{align}
  2nk - k^2 - 2k - (2n-k)\cdot 2 + 4
  &= 2nk - k^2 - 2k - 4n + 2k + 4 \nn\\
  &= 2n(k-2) - (k-2)(k+2) \nn\\
  &= (k-2)(2n-k-2).
\end{align}
(Momentum conservation contributes the $+4$.)  This is the amount of C integrals that is left to be done after using all the constraints on C and $\tilde C$.

\section{Useful Bessel function identities}\label{Bessel}
In this appendix we write few identities involving Bessel functions which are useful for the computation of the correlation function in the main text. 
\begin{align}
J_0(az)J_0(cz)
&=
\frac{1}{\pi}\int_0^\pi d\theta\,
J_0\!\left(z\sqrt{a^2+c^2-2ac\cos\theta}\right),
\\[0.5em]
\int_0^\infty dz\, z\, J_0(kz)\,K_0(bz)
&=
\frac{1}{k^2+b^2},
\\[0.5em]
\int_0^\pi \frac{d\theta}{A-B\cos\theta}
&=
\frac{\pi}{\sqrt{A^2-B^2}}.
\end{align}

\vspace{0.5em}

\begin{equation}
\int_0^\infty dz\, z\, J_0(|p_1|z)\,K_0(|p_2|z)\,J_0(|p_3|z)
=
\frac{1}{
\sqrt{
\left(|p_2|^2+(|p_1|+|p_3|)^2\right)
\left(|p_2|^2+(|p_1|-|p_3|)^2\right)
}
}.
\end{equation}


\bibliography{biblio}
\bibliographystyle{JHEP}
\end{document}